\documentclass[preprint,aps,amsmath,superscriptaddress,nofootinbib,tightenlines]{revtex4}
\usepackage{graphicx}
\usepackage{bm}
\usepackage{epsfig}

\newif\ifpdf
\ifx\pdfoutput\undefined
\pdffalse 
\else
\pdfoutput=1 
\pdftrue
\fi

\def\SppP{{\cal {P\!\!\!\!\hspace{0.04cm}\slash}}_\perp}

\def\nslash{n\!\!\!\slash}
\def\bnslash{\bar n\!\!\!\slash}
\def\pslash{p\!\!\!\slash}

\def\Aslash{A\!\!\!\slash}
\def\OMIT#1{}

\newcommand{\CH}[2]{\chi_{#1,#2}}
\newcommand{\CHp}[3]{\chi_{#1,#2}^{#3}}
\newcommand{\bCH}[2]{\overline\chi_{#1,#2}}
\newcommand{\bCHp}[3]{\overline\chi_{#1,#2}^{#3}}

\newcommand{\mfrac}[2]{\frac{\mbox{\normalsize $#1$}}{\mbox{\normalsize $#2$}} }
\newcommand{\nn}{\nonumber} 

\newcommand{\bn}{{\bar n}}
\newcommand{\bea}{\begin{eqnarray}}
\newcommand{\eea}{\end{eqnarray}}
\newcommand{\bnp}{\bar n \!\cdot\! p}
\newcommand{\bnP}{\bar {\cal P}}
\newcommand{\ppP}{{\cal P}_\perp}
\newcommand{\bnPd}{\bar {\cal P}^{\raisebox{0.8mm}{\scriptsize$\dagger$}} }
\newcommand{\cP}{{\cal P}}

\newcommand{\mcdot}{\!\cdot\!}

\newcommand{\cD}{{\cal D}}

\begin{document}
\ifpdf
\DeclareGraphicsExtensions{.pdf, .jpg}
\else
\DeclareGraphicsExtensions{.eps, .jpg}
\fi


\preprint{ \vbox{\hbox{UCSD/PTH 02-03} 
}}

\title{\phantom{x}\vspace{0.5cm} 
Hard Scattering Factorization from Effective Field Theory\\
\vspace{0.5cm} }

\author{Christian W.~Bauer}
\affiliation{Department of Physics, University of California at San Diego,
	La Jolla, CA 92093\footnote{Electronic address: bauer@einstein.ucsd.edu,
	pirjol@bose.ucsd.edu, iain@schwinger.ucsd.edu}}

\author{Sean Fleming}
\affiliation{Department of Physics, Carnegie Mellon University,
      	Pittsburgh, PA 15213\footnote{Electronic address: 
	fleming@kayenta.phys.cmu.edu, ira@cmuhep2.phys.cmu.edu}
	\vspace{0.7cm} }

\author{Dan Pirjol}
\affiliation{Department of Physics, University of California at San Diego,
	La Jolla, CA 92093\footnote{Electronic address: bauer@einstein.ucsd.edu, 
	pirjol@bose.ucsd.edu, iain@schwinger.ucsd.edu}}

\author{\\ Ira Z.~Rothstein}
\affiliation{Department of Physics, Carnegie Mellon University,
	Pittsburgh, PA 15213\footnote{Electronic address: 
	fleming@kayenta.phys.cmu.edu, ira@cmuhep2.phys.cmu.edu} 
	\vspace{0.7cm} }

\author{Iain W.~Stewart\vspace{0.5cm} }
\affiliation{Department of Physics, University of California at San Diego,
	La Jolla, CA 92093\footnote{Electronic address: bauer@einstein.ucsd.edu,
	pirjol@bose.ucsd.edu, iain@schwinger.ucsd.edu}}


\begin{abstract}
\vspace{0.5cm}
\setlength\baselineskip{18pt}

In this paper we show how gauge symmetries in an effective theory can be used to
simplify proofs of factorization formulae in highly energetic hadronic
processes. We use the soft-collinear effective theory, generalized to deal with
back-to-back jets of collinear particles. Our proofs do not depend on the choice
of a particular gauge, and the formalism is applicable to both exclusive and
inclusive factorization.  As examples we treat the $\pi$-$\gamma$ form factor
($\gamma\gamma^*\to \pi^0$), light meson form factors ($\gamma^* M \to M$), as
well as deep inelastic scattering ($e^- p\to e^- X$), Drell-Yan ($p\bar p\to X
\ell^+\ell^-$), and deeply virtual Compton scattering ($\gamma^* p \to
\gamma^{(*)} p$).

\end{abstract}

\maketitle


\newpage


\section{Introduction}

The principle of factorization underlies all theoretical predictions for
hadronic processes. Simply put, factorization is the statement that short and
long distances contributions to physical processes can be separated, up to
corrections suppressed by powers of the relevant large scale in the process. The
predictive power gained from this result stems from the fact that the
incalculable long distance effects are universal, defined in an unambiguous way
in terms of matrix elements. As a consequence, the non-perturbative long
distance effects can be extracted in one process and then used in another.  In
general, proving factorization is a difficult task~\cite{Ellis:ty}.  The proof
of factorization in Drell Yan processes, for instance, took several years to
sort out~\cite{DYrefs} (for reviews on factorization
see~\cite{Collins:1987pm,Sterman:1995fz,Jaffe:zw}). Indeed, there are still some
processes such as $B\to \pi\pi$ where a proof of factorization only exists at
one-loop~\cite{Bpipi}.

Given that we would like to retain our predictive power over the largest
possible range of energies, we are compelled to understand power corrections to
the factorized rates.  These corrections are not necessarily universal, and as
such, the relevant size of the power corrections are process dependent. In
processes for which there exists an operator product expansion (OPE), there is a
systematic way in which to include power corrections.  However, for a majority
of observables we do not have an OPE at our disposal, and the nature of the
power correction is not always known. For instance, in the case of shape
variables there is still some on going discussion about the form of subleading
corrections~\cite{nason,Korchemsky:1999kt}.

The purpose of this paper is to show that an effective theory framework can be
used to simplify proofs of factorization and describe processes with an operator
formalism. To do this we extend the soft-collinear effective theory (SCET)
developed in Refs.~\cite{bfl,bfps,cbis,bpssoft,bps}, to high energy processes.
It should be emphasized that there are several other useful advantages in using
an effective field theory (EFT).  For instance, the EFT makes any symmetries
which emerge in the $Q\to \infty$ limit manifest in the Lagrangian and
operators, and allow statements to be made to all orders in perturbation theory.
The calculation of hard coefficients reduces to simple matching calculations,
where subtracting the EFT graphs automatically removes all infrared divergences
from the QCD calculation.  Perhaps most importantly, it provides a framework for
systematically investigating power corrections.  Finally, the EFT framework
allows standard renormalization group techniques to be used for the resummation
of logarithms that are often necessary in calculating rates for certain high
energy scattering events~\cite{bfl,bfps,resum,resum2}.  The factorization
formulae that we prove in this paper are not new, but serve to illustrate our
approach in familiar settings. The results are valid to all orders in $\alpha_s$
and leading order in the power expansion.  The simplicity of our approach lies
in the fact that factorization occurs at the level of the SCET Lagrangian and
operators, and is facilitated by gauge symmetry in the EFT. This provides the
advantage that our proofs do not rely on making use of Ward identities and
induction, or on specifying a particular gauge.\footnote{In fact our
factorization proofs rely heavily on the gauge symmetry structure of SCET. When
a gauge fixing term is required for explicit calculations we use general
covariant gauges.}  Furthermore, it becomes rather simple to derive
factorization formulae for a myriad of processes, since many results are
universal. The examples given here serve to illustrate these simplifications.
Developments on the issues of power corrections and resummations are left to
future publications.\footnote{For recent work on subleading corrections in SCET
for heavy-to-light transitions see Ref.~\cite{Chay:2002vy}.}

In section~\ref{section_effective} we review the construction of the SCET. The
formalism developed in Refs.~\cite{bfl,bfps,cbis,bpssoft} is extended to include
two types of collinear particles moving in opposite directions in
section~\ref{section_nnbar}, and factorization for $\gamma^*$ to two collinear
states is discussed as an example. In section~\ref{section_melt} we define the
non-perturbative matrix elements such as the parton distribution functions that
will be needed for the processes presented in the paper, and in
section~\ref{section_symm} we discuss some of the symmetries in SCET that may be
used to place restrictions on matrix elements. In the remaining sections we give
various examples on how factorization theorems emerge in the effective theory
language. In section~\ref{section_exclusive} we prove factorization theorems for
two exclusive processes, namely the $\pi$-$\gamma$ form factor, and meson form
factors ($\gamma^* M \to M$) for arbitrary spin and isospin structure. In
section~\ref{section_inclusive} two inclusive processes are treated, namely DIS
($e^- p\to e^- X$) and Drell-Yan ($p\bar p\to X \ell^+\ell^-$), and we also give
results for deeply virtual Compton scattering ($\gamma^* p \to \gamma^{(*)}
p$). In these processes we include all leading power contributions in the
factorization proofs (even if the operators are only matched onto at higher
orders in perturbation theory such as for the gluon distribution functions). Our
conclusions are given in section~\ref{section_conclusion}.  In
appendix~\ref{app_fact} we show how auxiliary fields can be used to prove the
simultaneous factorization of soft fields from collinear fields for particles in
back-to-back directions.

\section{Formalism}\label{section_effective}

Effective field theories provide a simple and elegant way of organizing physics
in processes containing widely disparate energy scales. In constructing an EFT,
some degrees of freedom are eliminated, and the remaining degrees of freedom
must reproduce all the infrared physics of the full theory in the domain where
the EFT is valid. The EFT is organized by an expansion in $\lambda$, defined as
the ratio of small to large energy scales. As a useful guideline the following
steps are used to identify the infrared degrees of freedom: 1) Determine the
relevant scales in a problem from the size of the momenta and masses of all
particles that can make up the initial and final states, 2) Construct all
momenta from these scales whose components correspond to propagating degrees of
freedom, and which have offshellness less than the large scale,
i.e. $p^2-m^2\lesssim Q^2$.  Effective theory fields are then constructed for
each unique set of these momenta.

We will be interested in an EFT with particles of energy $Q$ much greater than
their mass.  The dynamics of these particles can be described by constructing a
soft-collinear effective theory (SCET). This theory is organized as an expansion
in powers of $\lambda\sim p_\perp/Q$, and offshell fluctuations with $p^2 \gg
(Q\lambda)^2$ are integrated out. In section~\ref{section_SCET} we begin by
describing this procedure and comparing the construction to other EFT's. We then
give a brief review of the soft-collinear effective theory developed in
Refs.~\cite{bfl,bfps,cbis,bpssoft,bps}. We do not attempt to give a
comprehensive treatment, but instead emphasize the main results and refer the
reader to the literature for details. In section \ref{section_nnbar} we extend
the formulation of SCET to describe processes with collinear particles moving in
back to back directions, and prove the factorization formula for $\gamma^*$ to
two collinear states as an example.  In section \ref{section_melt} we define the
non-perturbative matrix elements that are needed for our examples, then in
section~\ref{section_symm} we discuss some of the symmetry properties of
collinear fields and currents.

\subsection{Soft-Collinear Effective Theory} \label{section_SCET}

In the standard construction of an EFT one removes the short distance scales and
massive fields by integrating them out one at a time. A classical example is
integrating out the $W$ boson to obtain the effective electroweak Hamiltonian
with 4-fermion operators. However, in some situations we are interested in
integrating out large momentum fluctuations without fully removing the
corresponding field.  The simplest example of this is Heavy Quark Effective
Theory (HQET)~\cite{bbook}, which is constructed to describe the low energy
properties of mesons with a heavy quark.  Here the heavy anti-quarks are
integrated out and only heavy quarks with fluctuations close to their mass-shell
are retained.  This is accomplished by removing fluctuations of order the heavy
quark mass $m_Q$ with a field redefinition\cite{georgi}
\begin{equation} \label{fd1}
  \psi(x)=\sum_v e^{-im_Q v\cdot x}h_v(x) \,,
\end{equation}
where $v$ is the heavy quark velocity and $h_v$ is the field in the EFT. While
$\partial^\mu \psi(x)\sim m_Q\, \psi(x)$, the effective field has $\partial^\mu
h_v(x) \sim \Lambda_{\rm QCD}\, h_v(x)$, indicating that it no longer describes
short-distance fluctuations about the perturbative scale $m_Q$. Instead these
effects are encoded in calculable Wilson coefficients. The HQET degrees of
freedom with offshellness $p^2\sim \Lambda_{\rm QCD}^2$ are the heavy quarks,
soft gluons, and soft quarks.

Similarly, for collinear particles with energy $Q\gg m$, one needs to remove
momentum fluctuations $\sim\!Q$ while retaining effective theory fields to
describe smaller momenta. However, unlike heavy quarks the collinear particles
have two low energy scales. Consider the light-cone momenta, $p^+ = n\cdot p$
and $p^- = \bn\cdot p$ where $n^2=\bn^2=0$ and $n\cdot\bn=2$. Here $n$
parameterizes a light-cone direction close to that of the collinear particle and
$\bn$ the opposite direction (eg. for motion in the $z$ direction $n_\mu =
(1,0,0,1)$ and $\bn_\mu = (1,0,0,-1)$).  For a particle of mass $m\lesssim
p_\perp \ll Q$, we have $p^-\sim Q$, and a small parameter $\lambda \sim
p_\perp/Q$.  The scaling of the $p^+$ component is then fixed by the equations
of motion $p^+p^-+p^2_\perp =m^2$, so that $(p^+,p^-,p_\perp) \sim Q(\lambda^2,
1,\lambda)$.

The appearance of two small scales, $Q\lambda^2\ll Q\lambda \ll Q$, is similar
to the situation in non-relativistic QCD (NRQCD), which is an EFT for systems of
two heavy quarks with an expansion in their relative velocity $\beta$. In a
non-relativistic bound state the momentum of a heavy quark is ${\bf p}\sim
m_Q\beta$, but the equations of motion $E={\bf p}^2/(2 m_Q)$ make the energy
$E\sim m_Q\beta^2$, giving scales $m_Q\beta^2 \ll m_Q\beta\ll m_Q$. The two low
energy scales can be distinguished by following Eq.~(\ref{fd1}) with a further
field redefinition~\cite{LMR} $h_v(x) = \sum_{\bf p} e^{i{\bf p\cdot x}}
\psi_{\bf p}(x)$, so that derivatives on $\psi_{\bf p}$ only pick out the
$m\beta^2$ scale. The on-shell degrees of freedom are then the heavy quarks,
soft quarks and gluons with $p^2\sim (m_Q v)^2$, and ultrasoft quarks and gluons
with $p^2\sim (m_Q v^2)^2$.

\noindent{\bf\underline{SCET fields:}}\\ \indent For collinear particles the
analogous field redefinitions are~\cite{bfps,cbis}
\begin{equation}
  \phi(x)=\sum_n \sum_p e^{-ip\cdot x} \phi_{n,p}(x) \,,
\end{equation} 
where the collinear fields $\phi_{n,p}$ are labelled by light-cone vectors $n$
and label momentum $p$. Here $p$ contains the $\bn\cdot p\sim Q$ and
$p_\perp\sim Q\lambda$ momenta so that $\partial^\mu \phi_{n,p}\sim (Q\lambda^2)
\phi_{n,p}$. The field $\phi_{p}$ can either be a quark or gluon field. Similar
to $\psi_{\bf p}$, the missing $\sim Q$ fluctuations are described by Wilson
coefficients and the $\sim Q\lambda$ labels simplify the power counting by
distinguishing the $Q\lambda$ and $Q\lambda^2$ scales. Now
\begin{eqnarray} \label{phi+-}
  \phi_{n,p} \equiv \phi^+_{n,p} + \phi^-_{n,-p} \,,
\end{eqnarray}
so collinear particles and antiparticles are contained in the same effective
theory field, but have momentum labels with the opposite sign.  In the large
energy limit the four component fermion spinors contain two large and two small
components. One therefore defines collinear quark fields $\xi_{n,p}$ which only
retain the large components for motion in the $n$ direction and satisfies
$\nslash \xi_{n,p}=0$. For these fields $\xi_{n,p}^+/\xi_{n,p}^-$ destroy/create
the particles/antiparticles with large momentum $\bn\cdot p>0$ \cite{cbis}.  For
collinear gluons $A_{n,q}^{\mu\,\dagger}=A^\mu_{n,-q}$, and
$(A^\mu_{n,q})^+$/$(A^\mu_{n,q})^-$ destroy/create gluons with $\bn\mcdot q>0$.

For simplicity we will ignore quark masses and only consider massless $u$ and
$d$ quarks. For the processes considered here SCET then requires three types of
degrees of freedom: collinear, soft, and ultrasoft (usoft) fields. These are
distinguished by the scaling of the light cone components
($p^+$,$p^-$,$p^\perp$) of their momenta: $(\lambda^2,1,\lambda)$ for collinear
modes in the $n$ direction ($A_{n,q}$, $\xi_{n,p}$), $(\lambda,\lambda,\lambda)$
for the soft modes ($A_{q}^s$, $q_p^s$), and $(\lambda^2,\lambda^2,\lambda^2)$
for the usoft modes ($A_{us}$, $q_{us}$). The soft modes are labelled by their
order $Q\lambda$ momenta, so $A_{q}^s$ and $q_p^s$ are essentially just momentum
space fields. The usoft fields have no labels and depend only on the coordinate
$x$. The fields are assigned a scaling with $\lambda$ to make the action for
their kinetic terms order $\lambda^0$~\cite{bfl,bfps,bpssoft}. For instance
$\xi_{n,p}\sim\lambda$, $A_{n,q}^\mu \sim (\lambda^2,1,\lambda)$,
$A^s_q\sim\lambda$, and $A_{us}^\mu \sim \lambda^2$. At leading order only order
$\lambda^0$ vertices are necessary to correctly account for all order
$\lambda^0$ Feynman diagrams.

In HQET only external currents with momenta of order $m_b$ can change the label
$v$. Thus the Lagrangian has a superselection rule forbidding changes in the
four-velocity of the heavy quark~\cite{bbook,georgi}. In NRQCD the $v$ labels
are also conserved, but the smaller momentum labels ${\bf p}$ are changed by
operators in the effective theory such as the Coulomb potential.  A novel
feature of SCET is that interactions in the leading action can change both the
large and small parts of the momentum labels $p^\mu$. However, only external
currents can change the direction $n$ of a collinear particle, so this label is
conserved. Thus, for each distinct direction $n$ a separate set of collinear
fields are needed. In the remainder of this section we will restrict ourselves
to collinear particles with a single $n$. We will generalize the discussion to
the case of two back-to-back directions and discuss the factorization of
collinear particles with different $n$'s in section~\ref{section_nnbar}.

Since in SCET interactions can change the order $Q$ label momenta it turns out
to be very useful to introduce a label operator, ${\cal P}^\mu$~\cite{cbis}, for
which the collinear fields satisfy ${\cal P}^\mu \xi_{n,p} = p^\mu
\xi_{n,p}$. More generally, ${\cal P}^\mu$ acts on a product of labelled fields
as
\begin{eqnarray}
 f({\cal P}^\mu) \Big(\phi^\dagger_{q_1} \cdots \phi^\dagger_{q_m} 
 \phi_{p_1} \cdots \phi_{p_n}\Big) 
 = f( p_1^\mu\!+\!\ldots\!+\!p_n^\mu\! -\! q_1^\mu \!-\!\ldots\!-\! q_m^\mu)   
 \Big(\phi^\dagger_{q_1} \cdots \phi^\dagger_{q_m} \phi_{p_1} \cdots 
 \phi_{p_n}\Big) \,.
\end{eqnarray}
so conjugate field labels come with a minus sign. The operator $\cP_\mu$ acts to
the right, while the conjugate operator $\cP_\mu^\dagger$ acts to the left.  As
explained in Ref.~\cite{cbis} the label operator allows all large phases to be
moved to the front of operators with a factor $\exp({-ix\mcdot\cP})$. This phase
and the label sums can then be suppressed if we impose that interactions
conserve label momenta and that the momentum indices on fields are implicitly
summed over. Basically, for labels $p$ and $p'$ and residual momenta $k$ and
$k'$
\begin{eqnarray} \label{mcons}
  \int d^4x\ e^{i(p'-p+k'-k)\cdot x}
   = \delta(p-p') \int d^4x\ e^{i(k'-k)\cdot x} \,,
\end{eqnarray}
so that the label and residual momenta are individually conserved. (Although
technically the label momenta are discrete we abuse notation and use
$\delta(p\!-\!p')$ rather than $\delta_{p,p'}$ because it makes the subscripts
easier to read.) For convenience we define the operator $\bnP$ to pick out only
the order $\lambda^0$ labels on collinear fields, and the operator $\cP^\mu$ to
pick out only the order $\lambda$ labels.  For the matrix element of any
collinear operator ${\cal O}$, momentum conservation constrains the sum of field
labels~\cite{cbis}, giving
\begin{eqnarray} \label{MC}
  \big\langle M_{n,p_1} \big| \big[ f(\bnP)\, {\cal O} \big] 
    \big| M_{n,p_2}\big\rangle
   &=& f(\bn\mcdot(p_2\!-\!p_1))\ \big\langle M_{n,p_1}  \big|  
    {\cal O} \big| M_{n,p_2}\big\rangle \,,
\end{eqnarray}
for any function $f$.

For a single $n$ the Lagrangian can be broken up into three sectors: collinear,
usoft, and soft. We therefore write
\begin{eqnarray}
 {\cal L} = {\cal L}_{c,n}[\xi_{n,p},A_{n,q}^\mu,A_{us}^\mu] 
  + {\cal L}_{us}[q_{us},A^\mu_{us}] 
  + {\cal L}_{s}[q_{s,p}\,, A^\mu_{s,q}] \,, \label{sclag}
\end{eqnarray}
where we have made the field content of each sector explicit. We will 
discuss each of these terms separately.

\noindent{\bf\underline{Collinear sector:}}\\ \indent As explained in
detail in Ref.~\cite{bpssoft}, gauge invariance in SCET restricts the Lagrangian
and allowed form of operators.  Only local gauge transformations whose action is
closed on the effective theory fields need to be considered. These include
collinear, soft, and usoft transformations.  Each of these vary over different
distance scales, with collinear gauge transformations satisfying $\partial^\mu
U_n(x) \sim Q(\lambda^2, 1, \lambda)\: U_n(x)$, soft satisfying $\partial^\mu
V_{s}(x) \sim Q\lambda\: V_{s}(x)$, and usoft transformations with $\partial^\mu
V_{us}(x) \sim Q\lambda^2\: V_{us}(x)$. All particles transform under
$V_{us}(x)$ and usoft gluons act like background fields for collinear particles.
Invariance under $U_n(x)$ requires a collinear Wilson line built out of the
order $\lambda^0$ gluon fields~\cite{bfps,cbis}
\begin{eqnarray} \label{momW}
 W_n(x) = \bigg[
 \lower7pt\hbox{ $\stackrel{\mbox{\small$\sum$}}{\mbox{\footnotesize perms}}$ }
 {\rm exp} \Big( -\!g\, \frac{1}{\bnP}\, \bn \mcdot A_{n,q}(x)  \Big) \bigg] \,.
\end{eqnarray}
Here the operator $\bnP$ acts only inside the square brackets, the $n$ on $W_n$
refers to the direction of the collinear quanta, and $W_n$ is local with respect
to $x$ (corresponding to the residual momenta).  Taking the Fourier transform of
$\delta(\omega-\bnP) W_n(0)$ with respect to $\omega$ gives the more familiar
path-ordered Wilson line $W_n(y,-\infty)= {\rm P} \; {\rm exp}\big[ ig
\int^y_{-\infty} {\rm d}s\, \bn\mcdot A_n(s\bn)\big]$.  Under a collinear gauge
transformation $W_n$ transforms as $W_n\to U_n W_n$. An invariant
under collinear gauge transformations can therefore be formed by combining a
collinear fermion $\xi_{n,p}$ and the Wilson line $W_n$ in the form
\begin{eqnarray}\label{invariant}
  W_n^\dagger(x)\:  \xi_{n,p}(x)\,.
\end{eqnarray}
This combination still transforms under an usoft gauge transformation,
$W_n^\dagger\,\xi_{n,p}\to V_{us}(x)\: W_n^\dagger\,\xi_{n,p}$. We will often
suppress the $x$ dependence of the combination $W_n^\dagger\: \xi_{n,p}$.

Integrating out hard fluctuations gives Wilson coefficients in the effective
theory that are functions of the large $\bn\mcdot p_i$ collinear momenta,
$C(\bn\mcdot p_i)$. However, collinear gauge invariance restricts these
coefficients to only depend on the linear combination of momenta picked out by
the order $\lambda^0$ operator $\bnP$~\cite{cbis}. In general the Wilson
coefficients are then functions $C(\bnP,\bnP^\dagger)$ which must be inserted
between gauge invariant products of collinear fields.  In general the Wilson
coefficients also depend on the large momentum scales in a process such as $Q$
and the renormalization scale $\mu$.

To construct the collinear Lagrangian one can match full QCD onto operators with
collinear fields that are invariant under usoft and collinear gauge
transformations.  The collinear Lagrangian at order $\lambda^0$
is~\cite{bfps,cbis,bpssoft}
\begin{eqnarray} \label{Lc}
 {\cal L}_{c,n} 
 &=&   \bar\xi_{n,p'}\:  \bigg\{  i\, n\mcdot  D\!+\! g n\mcdot A_{n,q} 
  + \Big( \SppP\! + g \Aslash_{n,q}^\perp\Big)\, W\ \frac{1}{\bnP}\ W^\dagger\,
   \Big( \SppP\!  + g \Aslash_{n,q'}^\perp\Big) \bigg\}
  \frac{\bnslash}{2}\, \xi_{n,p}  \nn\\
 && + \frac{1}{2 g^2}\, {\rm tr}\ \bigg\{ 
    \Big[i\cD^\mu +g A_{n,q}^\mu \,, i\cD^\nu + g A_{n,q'}^\nu \Big]^{\,2} 
    \bigg\} + {\cal L}_c^{g.f.}\,,
\end{eqnarray}
where ${\cal L}_c^{g.f.}$ are gauge fixing terms, $iD^\mu = i\partial^\mu +
g A^\mu_{us}$, and 
\begin{eqnarray} \label{Dc}
  i\cD^\mu = \frac{n^\mu}{2}\, \bnP + {\cal P}_\perp^\mu + 
   \frac{\bn^\mu}{2}\, i\, n\mcdot D  \,.
\end{eqnarray}
Since usoft gluons act as background fields in the collinear gauge
transformation the couplings, $g(\mu)$, for both types of gluons must be
identical.

\noindent{\bf\underline{Usoft and Soft sectors:}}\\ \indent 
The usoft and soft Lagrangians for gluons and massless quarks are the same as
those in QCD. From Eq.~(\ref{sclag}) we see that collinear quarks and gluons
interact with usoft gluons, however at order $\lambda^0$ only the $n\mcdot
A_{us}$ component appears in Eq.~(\ref{Lc}).  In order to prove factorization
formulae it is essential to disentangle the collinear and usoft modes. This can
be done by introducing an usoft Wilson line
\begin{eqnarray} \label{Ydef}
 Y_n(x) &=&  {\rm P} \; {\rm exp}\bigg( ig \int^x_{-\infty}\!\!\! {\rm d}s\
   n\mcdot A_{us}(s n)\bigg) \,,
\end{eqnarray}
where the subscript $n$ on $Y_n$ labels the direction of the Wilson line (we
emphasize that this is different from the meaning of the subscript on $W_n$ in
Eq.~(\ref{momW})). An usoft gauge transformation takes $Y_n\to V_{us} Y_n$. In
Ref.~\cite{bpssoft} it was shown that the field redefinitions
\begin{eqnarray} \label{def0}
 \xi_{n,p} = Y_n\, \xi^{(0)}_{n,p} \,, \qquad\qquad
 A^\mu_{n,p} = Y_n\, A^{(0)\mu}_{n,p}\, Y^\dagger_n \,, 
\end{eqnarray}
imply $W_n = Y_n W^{(0)}_n Y^\dagger_n$ and decouple the usoft gluons from the
collinear particles in the leading order Lagrangian
\begin{eqnarray} \label{Lagrdecouple}
  {\cal L}_{c,n}[\xi_{n,p},A_{n,q}^\mu,n\mcdot A_{us}] 
    = {\cal L}_{c,n}[\xi_{n,p}^{(0)},A_{n,q}^{(0)\mu},0] \,.
\end{eqnarray}
Thus, the new collinear fields with superscript $(0)$ no longer interact with
usoft gluons or transform under an usoft gauge transformation.  Since the field
redefinitions do not change physical $S$ matrix elements, the new fields give an
equally valid parameterization of the collinear modes. The leading SCET
Lagrangian therefore factors into separate collinear and usoft sectors. This
alone does not guarantee factorization in operators and currents, since after
the field redefinition these operators may still contain both usoft and
collinear fields. However, the field redefinition makes factorization
transparent since identities such as $Y_n^\dagger Y_n=1$ may be applied directly
to the operators.  This will become clear in the examples in
sections~\ref{section_exclusive} and \ref{section_inclusive}.

The coupling of soft gluons to collinear particles differs from the
usoft-collinear interactions.  Interactions of a soft gluon with a collinear
particle results in a particle with momentum $p\sim Q(\lambda,1,\lambda)$, so
soft gluons can not appear in the collinear Lagrangian.  These offshell
particles have $p^2\sim Q^2\lambda$ and since $Q^2\lambda \gg (Q\lambda)^2$
these offshell quarks and gluons can be integrated out. At leading order in
$\lambda$ it was shown in Ref.~\cite{bpssoft} that in operators with collinear
fields this simply builds up factors of a soft Wilson line $S_n$ involving the
$n\cdot A_s$ component of the soft gluon field,
\begin{eqnarray} \label{Sdef}
 S_n &=& \bigg[ 
  \lower7pt \hbox{ $\stackrel{\mbox{\small$\sum$}}{\mbox{\small perms}}$ }
  \!\! \exp\bigg(-\!g\,\frac{1}{n\mcdot\cP}\ n\mcdot A_{s,q} \bigg) \bigg] \,.
\end{eqnarray}
The factors of $S_n$ appear outside gauge invariant products of collinear
fields, and their location is restricted by soft gauge invariance.

\subsection{SCET for $n$ and $\bn$ collinear fields}  \label{section_nnbar}

In this section we extend SCET to include the possibility of collinear fields
moving in different light-cone directions: $n_1$, $n_2$, $n_3, \dots$. These
directions can be considered to be distinct provided that $n_i\mcdot n_j\gg
\lambda^2$ for $i\ne j$. This follows from the fact that if $n_1\cdot n_2\sim
\lambda^2$ then the directions $n_1$ and $n_2$ are too close to be
distinguished. For example, a momentum $p_2=Q n_2$ can be considered to be
collinear in the $n_1$ direction if $n_1\cdot p_2 = Q n_1\cdot n_2 \sim
Q\lambda^2$, since this is the correct scaling for the small momentum component
of an $n_1$-collinear particle.

For simplicity we will only consider the case of back-to-back jets corresponding
to collinear particles moving in the $n$ and $\bn$ directions. These are clearly
distinct since $n\cdot\bn = 2$. Collinear particles in the $\bn$ direction have
$(+,-,\perp)$ momenta $\sim Q(1,\lambda^2,\lambda)$, and the $n\cdot p\sim 1$
and $p_\perp\sim \lambda$ momenta appear as labels on the corresponding fields:
$\xi_{\bn,p}$ and $A_{\bn,p}^\mu$. Emission of a collinear particle moving in
the $n$ direction from a collinear particle in the $\bn$ direction results in a
particle with momentum $k\sim Q(1,1,\lambda)$ and offshellness $k^2\sim
Q^2$. These offshell modes are integrated out to construct the SCET, so
collinear modes in the $n$ direction do not directly couple to collinear modes
in the $\bn$ direction.  A distinct set of collinear gauge transformations is
associated with each of $n$ and $\bn$, and fields in one direction do not
transform under the gauge symmetry associated with the opposite direction. Two
Wilson lines $W_n(x)$ and $W_\bn(x)$ are necessary (defined as in
Eq.~(\ref{momW})), and they appear in a way that makes collinear operators gauge
invariant. For instance the combinations
\begin{eqnarray} \label{invariant2}
 W_n^\dagger  \xi_{n,p}\,, \qquad W_\bn^\dagger  \xi_{\bn,p}
\end{eqnarray}
are invariant under collinear gauge transformations in the $n$ and $\bn$
directions, respectively.  We also require two types of label operators, $\bnP$
as before, and an operator $\cP$ to pick out $n\mcdot p$ labels that are order
$\lambda^0$.  Thus, $\bnP$ and $\cP$ act only on the $n$ and $\bn$ collinear
fields respectively. (The label operator $\cP^\mu$ still picks out order
$\lambda$ momentum components and therefore acts on both $n$ and $\bn$ fields.)
With two collinear directions, decoupling usoft gluons requires introducing both
$Y_n$ and $Y_\bn$ Wilson lines, defined as in Eq.~(\ref{Ydef}), but along the
$n$ or $\bn$ paths respectively. Finally, integrating out $\sim Q^2\lambda$
fluctuations at leading order induces both $S_n$ and $S_\bn$ soft Wilson lines
defined analogous to Eq.~(\ref{Sdef}). This is discussed in greater detail in
Appendix~\ref{app_fact} where we show explicitly to all orders in $g$ that
integrating out the $Q^2\lambda$ fluctuations causes
\begin{eqnarray}\label{Sadd}
  W_n^\dagger \xi_{n,p} \to S_{n} W_n^\dagger \xi_{n,p} \,,\qquad
  \bar\xi_{n,p} W_n  \to \bar\xi_{n,p} W_n  S_{n}^\dagger \,, \nn\\
  W_{\bn}^\dagger \xi_{\bn,p} \to S_{\bn} W_{\bn}^\dagger \xi_{\bn,p} \,,\qquad
  \bar\xi_{\bn,p} W_{\bn}  \to \bar\xi_{\bn,p} W_{\bn}  S_{\bn}^\dagger \,.
\end{eqnarray}
Relations for operators with collinear gluon fields are also derived in
Appendix~\ref{app_fact}.

Note that we have not included "Glauber gluons" with momenta $p^\mu\sim
(\lambda^2,\lambda^2,\lambda)$, which are kinematically allowed in $t$-channel
Coulomb exchange between $n$ and $\bn$ collinear quarks. In determining the
relevant degrees of freedom we have assumed that Glauber gluons are not
necessary to describe the infrared for the processes considered in this
paper. Intuitively, this can be seen from the fact these gluons are
instantaneous in both time and longitudinal separation, and only could
contribute when the $n$ and $\bn$ jets overlap for a duration of order
$1/(Q\lambda^2)$ in a space-time diagram. In processes with a hard interaction
the overlap scale is always much shorter than this (however this need not be the
case in processes such as forward scattering). For the Drell-Yan process more
quantitative arguments can be found in Refs.~\cite{DYrefs,CSSnew}.

At order $\lambda^0$ it is not possible to construct a gauge invariant kinetic
Lagrangian with terms that involve both $n$ and $\bn$ fields. Thus, the $n$ and
$\bn$ collinear modes are described by independent Lagrangians (however $n$ and
$\bn$ modes may still both appear in an external operator). The collinear
sector of the SCET Lagrangian is therefore
\begin{eqnarray}
 {\cal L}_{c,n}[\xi_{n,p},A_{n,q}^\mu,n\mcdot A_{us}]
 +{\cal L}_{c,\bn}[\xi_{\bn,p},A_{\bn,q}^\mu,\bn\mcdot A_{us}] \,.
\end{eqnarray}
Making the field redefinitions
\begin{eqnarray} \label{Yfd2}
 \xi_{n,p} &=& Y_n\, \xi^{(0)}_{n,p} \,, \qquad
 A^\mu_{n,p} = Y_n A^{(0)\mu}_{n,p} Y^\dagger_n \,, \nonumber \\
 \xi_{\bn,p} &=& Y_\bn\, \xi^{(0)}_{\bn,p} \,, \qquad
 A^\mu_{\bn,p} = Y_\bn A^{(0)\mu}_{\bn,p} Y^\dagger_\bn \,,
\end{eqnarray}
gives $W_n = Y_n W^{(0)}_n Y^\dagger_n$, $W_\bn = Y_\bn W^{(0)}_\bn
Y^\dagger_\bn$, and usoft degrees of freedom once again decouple from the
collinear modes since
\begin{eqnarray}
 {\cal L}_{c,n}[\xi_{n,p},A_{n,q}^\mu,n\mcdot A_{us}] \!+\! 
 {\cal L}_{c,\bn}[\xi_{\bn,p},A_{\bn,q}^\mu,\bn\mcdot A_{us}] \!=\! 
 {\cal L}_{c,n}[\xi_{n,p}^{(0)},A_{n,q}^{(0)\mu},0] \!+\! 
 {\cal L}_{c,\bn}[\xi_{\bn,p}^{(0)},A_{\bn,q}^{(0)\mu},0]. 
\end{eqnarray}
Thus, usoft gluons are removed from the collinear Lagrangian at the expense of
inducing $Y_n$ and $Y_\bn$ factors in operators with collinear fields. In
certain cases the identities $Y_n^\dagger Y_n=1$ and $Y_\bn^\dagger Y_\bn=1$ can
be used in these operators to cancel usoft gluon interactions. Perturbatively
these cancellations would occur by adding an infinite set of Feynman diagrams.

To see in more detail how this works consider the simple example of the
$\gamma^*$-production of back-to-back collinear states $X_n$ and $X_\bn$.  The
full theory current $\bar\psi(x) \Gamma \psi(x)$ matches onto an effective theory
operator $O_{n\bn}$. Naively one might guess that the SCET operator mediating
this process is
\begin{eqnarray}
  O_{n\bn} = \bar \xi_{n,p_1} \Gamma \xi_{\bn, p_2}\,.
\end{eqnarray}
However, this operator is not invariant under the collinear gauge
transformations ${U}_{n}$ and ${U}_{\bn}$, so the process is instead mediated by
the invariant operator
\begin{eqnarray}
  O_{n\bn} = \bar \xi_{n,p_1} W_n\, \Gamma\, W^\dagger_\bn\, \xi_{\bn, p_2}\,.
\end{eqnarray}
A hard matching coefficient $C(\bnP, \bnP^\dagger, \cP, \cP^\dagger)$ can be
inserted in any location in the operator that does not break apart the gauge
invariant combinations of fields in Eq.~(\ref{invariant2}). The operators $\bnP$
and $\cP$ in the coefficient only pick out momenta that are order $\lambda^0$ in
the power counting. Thus, $\cP$ does not act on fields in the $n$ direction and
$\bnP$ does not act on fields in the $\bn$ direction, and the most general
result is
\begin{eqnarray}
 O_{n\bn} = \bar \xi_{n,p_1} W_n\, \Gamma\, C(\bnP^\dagger, \cP) 
 W^\dagger_\bn\, \xi_{\bn, p_2}\,.
\end{eqnarray}
Next, we integrate out the offshell $Q^2\lambda$ fluctuations which induces
additional soft Wilson lines in the operator. This is discussed in detail in
Appendix~\ref{app_fact} and from Eq.~(\ref{Sadd}) gives
\begin{eqnarray} \label{OnnWS}
 O_{n\bn} = \bar \xi_{n,p_1} W_n S_n^\dagger \, \Gamma\, 
    C(\bnP^\dagger, \cP) S_\bn \, W^\dagger_\bn\, \xi_{\bn, p_2}\,.
\end{eqnarray}
Note that $\bnP$ and $\cP$ do not act on the fields in the soft Wilson lines
since soft gluons carry only order $\lambda$ momenta.  Finally, we can make the
usoft gluon couplings explicit by switching to the $(0)$ fields using
Eq.~(\ref{Yfd2}):
\begin{eqnarray}\label{exop}
 O_{n\bn} = \bar \xi^{(0)}_{n,p_1} W^{(0)}_n Y^\dagger_n S_n^\dagger \, \Gamma\,
 C(\bnP^\dagger, \cP)\, S_\bn Y_\bn  W^{(0)\dagger}_\bn\xi^{(0)}_{\bn, p_2}\,. 
\end{eqnarray}
This operator is manifestly invariant under collinear gauge transformation in
the $n$ and $\bn$ direction, as well as under soft and usoft gauge
transformations.  

To separate the short distance Wilson coefficient from the long-distance 
operator one introduces convolution variables $\omega$ and $\omega'$ to give
\begin{eqnarray}\label{Onnf}
 O_{n\bn}&=& \int\!\!{d\omega\, d\omega'} C(\omega,\omega') 
 O_{n\bn}(\omega,\omega') \,,\\
 O_{n\bn}(\omega,\omega') &=& \Big[ \bar
 \xi_{n,p_1}^{(0)} W_n^{(0)} \delta(\bnP^\dagger\!-\!\omega)  
 Y^\dagger_n S^\dagger_n \, \Gamma\, S_\bn Y_\bn \delta(\cP\!-\!\omega') 
 W^{\dagger(0)}_\bn \xi_{\bn, p_2}^{(0)} \Big] \,. \nn
\end{eqnarray}
The function $C(\omega,\omega')$ contains all the short distance physics and is
determined by matching the full theory onto this effective theory
operator. $O_{n\bn}(\omega,\omega')$ contains all the infrared long-distance
QCD contributions at leading order in $\lambda$.

Now consider the matrix element of the production current between $\langle X_n
X_\bn |$ and the vacuum.  Taking the $\gamma^*$ to have large time-like momentum
$q^\mu=(Q,0,0,0)$ (and zero residual momentum) we have
\begin{eqnarray} \label{qqff}
 && \int\!\! d^4x\, e^{-i q\cdot x}\, \big\langle X_n X_\bn \big| \bar\psi(x)
 \Gamma \psi(x) \big| 0 \big\rangle = 
 \int\!\! d^4x\, \big\langle X_n X_\bn\big| O_{n\bn}(x) 
 \big| 0 \big\rangle \nn \\
 && \quad = \int\!\! d^4x\, e^{ik\cdot x}\, \big\langle (X_n X_\bn)(k)
 \big| O_{n\bn}(x=0) \big| 0 \big\rangle 
 = \big\langle (X_n X_\bn)(0)\big| O_{n\bn}(0)  \big| 0 \big\rangle \,.
\end{eqnarray}
In the first step the conservation of the large label momentum $q$ was made
implicit in the matrix element (c.f. Eq.~(\ref{mcons})). Since we are in the
center-of-mass frame the $X_n$ has large momentum $\bn\cdot p=Q$, and the $X_\bn$
has momentum $n\cdot p'=Q$.  Now using translation invariance, we see that the
remaining $x$ integral forces the $|X_n X_\bn\rangle$ state to have zero
residual momentum.  Using Eq.~(\ref{Onnf}) this matrix element is equal to
\begin{eqnarray}
  \int\!\!{d\omega\, d\omega'} C(\omega,\omega') \big\langle X_n\bar X_\bn\big|
    O_{n\bn}(\omega,\omega') \big| 0 \big\rangle 
 =\int\!\!{d\omega\, d\omega'}\, C(\omega,\omega')\: J_n(\omega)\:
  \Gamma\, S_{n\bn}  \: J_\bn(\omega') \,,\ \ 
\end{eqnarray}
where the functions $C$, $J_n$, $J_\bn$, and $S$ also depend on the
renormalization point $\mu$. Here we have used the fact that both $X_n$ and
$X_\bn$ can be described entirely by collinear particles in the $n$ and $\bn$
directions respectively.  Since the Lagrangians for the collinear, soft, and
usoft fields are factorized the remaining matrix element splits into distinct
matrix elements for each class of modes. These matrix elements are
\begin{eqnarray}  \label{JJSY}
 J_n(\omega) &=& \big\langle X_n(Q/2) \big| \bar\xi_{n,p_1}^{(0)} W_n^{(0)} 
     \delta(\bnP^\dagger\!-\!\omega) \big| 0\rangle   \,,\nn\\
 J_\bn(\omega') &=& \big\langle X_\bn(Q/2) \big| \delta(\cP\!-\!\omega')  
     W^{\dagger(0)}_\bn  \xi_{\bn, p_2}^{(0)} \big| 0\rangle   \,,\nn\\
 S_{n\bn} &=& \big\langle 0 \big |  Y_n^\dagger S_n^\dagger S_\bn Y_\bn 
   \big| 0 \big\rangle \,,
\end{eqnarray}
(and are matrices whose color, spin, and flavor indices are suppressed). Note
that $J_n$, $J_\bn$, and $S_{n\bn}$ are explicitly invariant under the
collinear, soft, and usoft gauge transformations~\cite{bpssoft} of SCET, but
still transform globally under a color rotation.  Now using the momentum
conservation identity in Eq.~(\ref{MC}), the large momentum of the $X_n$ and
$X_\bn$ states set $\cP\to Q$ and $\bnP^\dagger\to Q$. Label conservation also
implies that the total perpendicular momentum of each of $J_n$, $J_\bn$, and
$S_{n\bn}$ is zero.  The sum over $\omega$ and $\omega'$ can then be performed
to give the final factorized form
\begin{eqnarray} \label{jjrslt}
  C(Q,Q)\: J_n(Q)\: \Gamma\, S_{n\bn}  \: J_\bn(Q) \,.
\end{eqnarray}

Although rather idealized, the above example illustrates the main steps needed
to derive a factorization formula.  Taking $X_n$ and $X_\bn$ to be single quark
states the result in Eq.~(\ref{jjrslt}) also agrees with the factorization
formula for the onshell production form factor for $q\bar q$
\cite{Korchemskyff,CollinsP}.\footnote{In this case depending on the choice of
infrared regulator(s), it may not be possible to distinguish the $Y_n$ and $S_n$
Wilson lines in $S_{n\bn}(\mu)$. For instance if one chooses $\Lambda_{\rm
IR}\sim Q\lambda$ then the usoft gluons give scaleless loop integrals and can be
dropped, so that $Y_n^\dagger S_n^\dagger S_\bn Y_\bn\to S_n^\dagger S_\bn$. If
instead one chooses $\Lambda_{\rm IR}\sim Q\lambda^2$ then the soft gluons give
scaleless loop integrals (they simply act to pull-up the ultraviolet divergences
in the usoft integrals to the hard scale~\cite{amis3,hms1}), so the soft Wilson
lines can be suppressed. This is why one only finds $S_n^\dagger S_\bn$ for this
operator in the literature. For typical regulator choices the other gluons are
simply not required to reproduce the infrared structure of the full theory
result.}  In the above example the factors of $S$ and $Y$ in the operator in
Eq.~(\ref{exop}) do not cancel. In the examples we will consider in
sections~\ref{section_exclusive} and \ref{section_inclusive} there are several
operators at leading order, however the factors of $S$ and $Y$ cancel in
observable matrix elements. The collinear matrix elements of long-distance
operators, such as those in Eq.~(\ref{JJSY}), are the ones that have
interpretations as structure functions or wave functions. In the next section we
give the operator definitions for these functions that will be needed in the
remainder of the paper.

\subsection{Non-perturbative Matrix Elements} \label{section_melt}

Predictions for hadronic processes depend on universal matrix elements that are
not computable in perturbation theory. For exclusive processes these include
light-cone wavefunctions and form factors, while for inclusive processes they
include parton distribution functions and fragmentation functions. In this
section we define matrix elements in SCET that are needed for our examples.  All
the collinear operators considered here decouple from usoft gluons since they
are local in the residual coordinate $x$ and because $Y^\dagger(x) Y(x)=1$. Thus
\begin{eqnarray} \label{decoup}
  \bar\xi_{n,p'} W\Gamma W^\dagger \xi_{n,p} = 
  \bar\xi_{n,p'}^{(0)} W^{(0)}\Gamma W^{(0)\dagger} \xi_{n,p}^{(0)}\,,
\end{eqnarray}
and expressions with and without the $(0)$ superscript are equal.  For
convenience we will write the collinear fields without the superscript in the
remainder of this section.

Consider first the light-cone wavefunctions. For the pion isotriplet $\pi^a$,
the wavefunction $\phi_{\pi}(x)$ is conventionally defined by \cite{brodskylepage}
\begin{eqnarray} \label{piwfn_rest}
 \big\langle \pi^a(p)\big|\, \bar{\psi}(y)\gamma^\mu \gamma^5 
  \frac{\tau^b}{\sqrt{2}} \,Y(y,x)\, \psi(x)\, \big| 0 \big\rangle
  = -i \OMIT{\sqrt{2}\,} f_\pi \delta^{ab} p^\mu\! \int_0^1\!\! dz\, 
  e^{i[z p\cdot y+(1-z)p \cdot x]}\, \phi_\pi(\mu,z),
\end{eqnarray}
Here $f_\pi\simeq 131\,{\rm MeV}$ and the QCD field $\psi$ denotes the isospin
doublet $\{\psi^{(u)}, \psi^{(d)}\}$. The coordinates satisfy $(y-x)^2=0$, which
ensures that the path from $y^\mu$ to $x^\mu$ is along the light-cone, and
$Y(y,x)$ is a Wilson line along this path. In SCET we require the matrix
elements of highly energetic pions, which therefore have collinear
constituents. Boosting the matrix element in Eq.~(\ref{piwfn_rest}) and letting
$y^\mu=y \bn^\mu$, $x^\mu=x \bn^\mu$ we have
\begin{eqnarray} \label{piwfn_collin}
 \big\langle \pi^a_{n,p}\big|\, \bar{\xi}_{n,y}\, 
 \Gamma_\pi^b\, W(y,x)\, \xi_{n,x}\, \big| 0  \big\rangle
  = -if_\pi\, {\bn\mcdot p}\,\delta^{ab} \,\int_0^1\!\! dz\, 
  e^{i \bn\cdot p [z y+(1-z) x]}\, \phi_\pi(\mu,z),
\end{eqnarray}
where $\Gamma_\pi^b={\bnslash\gamma_5\tau^b}/{\sqrt{2}}$, and $\xi_{n,z}$ is a
collinear field with position space label $z\bn^\mu$.  For our purposes it is
more useful to use the operator with momentum space labels~\cite{bps}
\begin{eqnarray} \label{pimom}
  \big\langle \pi_{n,p}^a \big| \bar{\xi}_{n,p_1} W \Gamma_\pi^b 
  \delta(\omega\!-\!\bnP_+) W^\dagger \xi_{n,p_2}\big| 0
  \big\rangle
 &=&\int \frac{dy}{2 \pi}\, e^{-i\omega y} \big\langle \pi_{n,p}^a \big| 
  \bar{\xi}_{n,y} \Gamma_\pi^b W(y,-y) \xi_{n,-y} \big| 0
  \big\rangle \\
 &=&\!\! -{i}f_\pi\, {\bn\mcdot p}\,\delta^{ab}\!\! 
  \int_0^1\!\! dz\, 
  \delta[\omega\!-\!(2z\!-\!1)\bn\mcdot p]\, \phi_\pi(\mu,z)\,, \nn
\end{eqnarray}
where $\bnP_\pm=\bnP^\dagger\pm\bnP$. 
\OMIT{ 
Similarly, the light-cone wavefunction 
of the longitudinal $\rho$ is defined by the matrix element
\begin{eqnarray} \label{rhomom}
\big\langle \rho^a_{n,p}(\epsilon)\big| \bar\xi_{n,p_1} W \Gamma_{\parallel}^b
  \delta(\omega\!-\!\bnP_+) W^\dagger \xi_{n,p_2} \big| 0\rangle \nn\\
  = - {i} f_\rho m_\rho(\bn\mcdot\epsilon^*)\delta^{ab}
   \!\!\!\!\!\!&& 
   \int_0^1\!\!\! dz\,\delta[\omega\!-\!(2z\!-\!1)\bn\mcdot p] 
    \phi_\rho^\parallel(\mu,z)\,,
\OMIT{\nn\\
\big\langle \rho^a_{n,p}(\epsilon)\big| \bar\xi_{n,p_1}W \Gamma_{\perp}^b
  \delta(\omega\!-\!\bnP^+) W^\dagger \xi_{n,p_2} \big| 0\rangle 
  &=& - 2 if^{}_{\!\perp}\bn\mcdot p\, \epsilon^{*\mu}_\perp\delta^{ab}\!  
    \int_0^1\!\!\!  dz\, \delta[\omega\!-\!(2z\!-\!1)\bn\mcdot p] 
    \phi_\rho^\perp(z)\,,}
\end{eqnarray}
where $\Gamma_{\parallel}^b= {\bnslash\tau^b}/{\sqrt{2}}$. The light-cone
wavefunction of a transverse $\rho$ will not be used.  } 
In Eq.~(\ref{pimom}) the delta function fixes $\omega$ to the sum of labels
picked out by the $\bnP_+$ operator. The combination picked out by $\bnP_-$ is
equivalent to ($-\bnP$) acting on the entire operator, and using Eq.~(\ref{MC})
is fixed to the $\bn\mcdot p$ momentum of the pion state.

In some situations it is convenient to have delta functions which fix the labels
of both $W^\dagger\xi_{n,p}$ and $\bar\xi_{n,p}W$. In this case a useful field
is
\begin{eqnarray} \label{CH1}
  \CH{n}{\omega}^{\,(i)} &\equiv& \big[ \delta(\omega-\bnP)\, 
     W_n^\dagger\xi_{n,p}^{(i)} \big] \,.
\end{eqnarray}
Here $i$ is the flavor index and will be omitted if the flavor doublet field is
implied. 
\OMIT{If a fixed perpendicular momentum is also needed we will use the
notation
\begin{eqnarray} \label{CH2}
  \CHp{n}{\omega}{\,\omega_{\!\perp}} &\equiv& \big[\delta(\omega-\bnP)
     \delta(\omega_\perp-\ppP)\, W_n^\dagger\xi_{n,p} \big] \,.
\end{eqnarray}
} Note that unlike the $p$ in $\xi_{n,p}$ the label $\omega$ on
$\CHp{n}{\omega}{}$ is not summed over.  A matrix element with $\chi_{n,\omega}$
fields is related to a matrix element like the one in Eq.~(\ref{pimom}) through
\begin{eqnarray} \label{pimom2}
 \big\langle M_{n,p} \big| \bCH{n}{\omega}\Gamma
 \CH{n}{\omega'} \big| M_{n,p'} \big\rangle
 &=& 2\, \delta(\omega_-\!-\!\bn\mcdot p_-)\, \big\langle M_{n,p} \big| 
  \bar{\xi}_{n,p_1}W \Gamma  \delta(\omega_+\!-\!\bnP_+) 
  W^\dagger \xi_{n,p_2}\big| M_{n,p'} \big\rangle \,,\nn\\
\end{eqnarray}
where $\omega_\pm = \omega\pm\omega'$ and $p_-=p-p'$. Thus, with the $\chi$
notation the momentum conserving delta functions become explicit. The factor of
two appears from treating the $\omega$'s as continuous variables, and in the
final results cancels with a factor of $1/2$ from a Jacobian.

For inclusive processes such as DIS it is the proton parton distribution
functions for quarks of flavor $i$, $f_{i/p}(z)$, antiquarks $\bar f_{i/p}(z)$,
and gluons, $f_{g/p}(z)$ which are needed.  The standard coordinate space
definitions~\cite{Soper} are ($y^\mu=y \bn^\mu$)
\begin{eqnarray}
 f_{i/p}(z) &=& \int\!\!\frac{dy}{2\pi}\: e^{-i\, 2z\bn\cdot p\, y}
  \big\langle p \big| \,\bar\psi^{(i)}(y) Y(y,-y) \,{\bnslash}\,
  \psi^{(i)}(-y)\big| p \big\rangle \Big|_{\mbox{\footnotesize spin avg.}} \,,\\
 f_{g/p}(z) &=& \frac{2}{z\: \bn\mcdot p} \int\!\!\frac{dy}{2\pi}\:
  e^{-i\, 2z\bn\cdot p\, y}\, \bn_\mu\bn^\nu
  \big\langle p \big| \,G_a^{\mu\lambda}(y) Y^{ab}(y,-y) 
  G^b_{\lambda\nu}(-y)\big| p \big\rangle\Big|_{\mbox{\footnotesize spin avg.}}
  \,, \nn
\end{eqnarray}
and $\bar f_{i/p}(z)=-f_{i/p}(-z)$.  Here $G^a_{\mu\lambda}(y)$ is the gluon
field strength, $Y(y,-y)$ and $Y^{ab}(y,-y)$ are path-ordered Wilson lines in
the fundamental and adjoint representations, and $|p\rangle$ is the proton state
with momentum $p$. In SCET these distribution functions can be defined by the
matrix elements of collinear fields with collinear proton states
\begin{eqnarray} \label{pdfqg}
 \frac12\sum_{\rm spin}  
 \big\langle p_{n} \big| \bCH{n}{\omega}^{\,(i)} \:{\bnslash}\:
   \CH{n}{\omega'}^{\,(i)} \big| p_{n} \big\rangle 
   &=&
   4\bn\mcdot p \int_0^1\!\!\! dz \,\delta(\omega_-)
   \delta(\omega_+ \!-\! 2z \, \bn\mcdot p)\, f_{i/p}(z)  \\
   &-&
   4\bn\mcdot p \int_0^1\!\!\! dz \,\delta(\omega_-)
   \delta(\omega_+ \!+\! 2z \, \bn\mcdot p)\, \bar f_{i/p}(z)\,, \nn \\
  \frac12\sum_{\rm spin} \big\langle p_{n} \big| {\rm Tr}\:\big[  
  B^\mu_{n,\omega} B_\mu^{n,\omega'} \big] \big| p_{n} \big\rangle \!
   &=& -\frac{\omega_+ \,\bn\mcdot p}{2} \int_0^1\!\!\! dz
  \,\delta(\omega_-)\delta(\omega_+ \!-\!2 z \, \bn\mcdot p)  f_{g/p}(z) \nn ,
\end{eqnarray}
where $\omega_\pm=\omega\pm\omega'$, and $B^\mu_{n,\omega}\equiv\bn_\nu ({\cal
G}_{n,\omega})^{\nu\mu}$ with the collinear gauge invariant field strength 
\begin{eqnarray} \label{covG}
  ({\cal G}_{n,\omega})^{\mu\nu} = -\frac{i}{g}\Big[ \delta(\omega-\bnP)
   W^\dagger [i{\cal D}_n^\mu + gA_{n,q}^\mu, i{\cal D}_n^\nu+gA_{n,q'}^\nu ] W
   \Big] \,.
\end{eqnarray}
Both operators in Eq.~(\ref{pdfqg}) are order $\lambda^2$ since
$\xi_{n,\omega}\sim B^{\perp}_{n,\omega} \sim \lambda$. Note that the matrix
element of a single operator ($\bCH{n}{\omega}^{\,(i)} \:{\bnslash}\:
\CH{n}{\omega'}^{\,(i)}$) contains both the quark and antiquark distributions.
This is due to Eq.~(\ref{phi+-}), from which we see that for $\omega=\omega'>0$
($\omega=\omega'<0$) this operator reduces to the number operator for collinear
quarks (antiquarks) with momentum $\omega$.

Processes other than DIS sometimes depend on more complicated distribution
functions.  
\OMIT{For Drell-Yan the distribution functions that depend on
perpendicular momenta appear at intermediate stages, and for quarks these are
defined by 
\begin{eqnarray} \label{genpdf}
  \frac12\sum_{\rm spin}
  \big\langle p_{n,p} \big| \bCHp{n}{\omega}{\:\omega_{\!\perp}}  \,
  {\bnslash}\, \CHp{n}{\omega'}{\,\omega_{\!\perp}'} 
  \big| p_{n,p} \big\rangle 
   &=&  4\bn\mcdot p \int_0^1\!\!\! dz \,\delta(\omega_-)\delta(\omega^\perp_-)
   \delta(\omega_+ \!-\! 2z \, \bn\mcdot p) \tilde f_{i/p}(z,\omega^\perp_+)\\
  &-& 4\bn\mcdot p \int_0^1\!\!\! dz \,\delta(\omega_-)\delta(\omega^\perp_-)
   \delta(\omega_+ \!+\! 2z \, \bn\mcdot p) {\tilde{\bar f}}_{i/p}(z, 
   \omega^\perp_+) \,,
\end{eqnarray}
where $\omega^\perp_\pm=\omega_\perp\pm \omega_\perp'$. The gluon distribution
$\tilde f_{g/p}(z,\omega_+^\perp)$ can be defined in a similar way.}  In deeply
virtual Compton scattering (DVCS) we will need to parameterize the matrix
element of an operator between proton states with different momenta.  In terms of
QCD fields $\psi$ the nonforward parton distribution function (NFPDF) 
defined by Radyushkin in Eq.(4.1) of Ref.~\cite{Radyushkin:1997ki} are (up to a 
trivial translation)
\begin{eqnarray}\label{asyqpdf}
& & \big\langle p^{\prime}, \sigma^\prime \big| 
  \bar\psi^{(i)}(y) Y(y,-y) \,{\bnslash}\, \psi^{(i)}(-y)
  \big| p,\sigma \big\rangle \\ 
 &&\qquad = \, e(\sigma^\prime,\sigma) 
  \int^1_0 dz \big[ e^{i\bn \cdot p(2z-\zeta)y}\: {\cal F}^{(i)}_\zeta(z;t)
  -e^{-i\bn\cdot p(2z-\zeta)y}\: \overline{{\cal F}}^{(i)}_\zeta(z;t) \big]\nn\\
&& \qquad\ + \, h(\sigma^\prime,\sigma) 
  \int^1_0 dz \big[ e^{i\bn \cdot p(2z-\zeta)y}\: {\cal K}^{(i)}_\zeta(z;t)  
  - e^{-i\bn\cdot p(2z-\zeta)y}\: \overline{{\cal K}}^{(i)}_\zeta(z;t)
  \big] \,,\nn
\end{eqnarray}
where $t = (p-p^\prime)^2$, and $\zeta = 1-\bnp^\prime/\bnp$. Here
$e(\sigma,\sigma^\prime)$ and $h(\sigma,\sigma^\prime)$ are matrix elements
which respectively preserve or flip the proton spin. They are defined in terms
of the proton spinors
\begin{eqnarray}
e(\sigma^\prime,\sigma) &=&  
 \bar{u}(p',\sigma')\: {\bnslash}\: u(p,\sigma)
  \,,\qquad
h(\sigma^\prime,\sigma) = \frac{1}{2m_p}\,  
 \bar{u}(p',\sigma')\: {[\bnslash, \pslash\!-\!\pslash']}\: 
 u(p,\sigma) \,, 
\end{eqnarray}
where $m_p$ is the proton mass. The NFPDF for gluons is similarly given
by~\cite{Radyushkin:1997ki}
\begin{eqnarray}
& & \bn_\mu \bn^\nu \big\langle p^{\prime}, \sigma^\prime \big| 
  \,G_a^{\mu\lambda}(y) Y^{ab}(y,-y) G^b_{\lambda\nu}(-y)
  \big| p,\sigma \big\rangle  \\
&&  \qquad\qquad  = \frac{\bn\mcdot p}{2}\: e(\sigma^\prime,\sigma)  
  \int^1_0 dz \left[ e^{i\bn\cdot p(2z-\zeta)y} + e^{-i\bn\cdot p(2z-\zeta)y} 
 \right] {\cal F}^{g}_\zeta(z;t) \nonumber \\
&& \qquad\qquad \ +  \frac{\bn\mcdot p}{2}\: h(\sigma^\prime,\sigma) 
  \int^1_0 dz \left[ e^{i\bn\cdot p(2z-\zeta)y} 
  + e^{-i\bn\cdot p(2z-\zeta)y}\right] {\cal K}^{g}_\zeta(z;t) \,. \nn
\end{eqnarray}
In SCET the definition of the NFPDFs in terms of collinear fields are
\begin{eqnarray} \label{nfpdfqg}
 & &  \big\langle p_n^\prime,\sigma^\prime \big| 
  \bCH{n}{\omega}^{\,(i)} \:{\bnslash}\: \CH{n}{\omega'}^{\,(i)} 
  \big| p_n, \sigma \big\rangle = 2\delta(\omega_-+\bn\mcdot p\,\zeta) 
  \int_0^1\!\!\! dz \, \\
&& \qquad\quad
    \times\Big\{  
   e(\sigma',\sigma) \big[ 
   \delta(\omega_+ \!-\! (2z\!-\!\zeta) \, \bn\mcdot p)\, 
   {\cal F}^{(i)}_\zeta(z;t)    -  \delta(\omega_+ \!+\!
   (2z\!-\!\zeta) \, \bn\mcdot p)\, \overline{{\cal F}}^{(i)}_\zeta(z;t) 
   \big] 
\nonumber \\
&& \qquad\quad
  + h(\sigma^\prime,\sigma)  \big[ 
   \delta(\omega_+ \!-\! (2z\!-\!\zeta) \, \bn\mcdot p)\, {\cal
   K}^{(i)}_\zeta(z;t)-  \delta(\omega_+ \!+\! (2z\!-\!\zeta) \, 
  \bn\mcdot p)\, \overline{{\cal K}}^{(i)}_\zeta(z;t) \big] \Big\}
\,, \nn \\
&&  \big\langle p_{n}',\sigma' \big| {\rm Tr}\:\big[ 
  B^\mu_{n,\omega} B_\mu^{n,\omega'}  \big] \big| p_{n},\sigma \big\rangle \! = 
  -\frac{\bn\mcdot p}{2}\: \delta(\omega_-+\bn\mcdot p\, \zeta) 
  \int_0^1\!\!\! dz \nonumber \\
 && \qquad\quad 
  \times \Big\{ e(\sigma^\prime,\sigma) 
  \,\delta(\omega_+ \!-\!(2z\!-\!\zeta)
  \, \bn\mcdot p) {\cal F}^{g}_\zeta(z;t) 
  + h(\sigma^\prime,\sigma)   \delta(\omega_+ \!-\!(2z\!-\!\zeta)\,\bn\mcdot p) 
  {\cal K}^{g}_\zeta(z,t)  \Big\} \,,\nn
\end{eqnarray}
where the spinors in $e(\sigma',\sigma)$ and $h(\sigma',\sigma)$ are two 
component effective theory spinors, so that $u\to u_n$ where $\nslash u_n=0$. 
Note that for $p^\prime \to p$ both $\zeta \to 0$ and $ t \to 0$. In
this limit the NFPDFs reduce to the standard PDFs:
\begin{eqnarray}
\lim_{p \to p^\prime} {\cal F}^{(i)}_\zeta &=& f_{i/p}(z)
\,,\qquad
\lim_{p \to p^\prime} \overline{{\cal F}}^{(i)}_\zeta =
\bar{f}_{i/p}(z)
\,, \qquad
\lim_{p \to p^\prime} {\cal F}^{(g)}_\zeta = z\, f_{g/p}(z) \,.
\end{eqnarray}

\subsection{Symmetries for collinear fields} \label{section_symm}

In this section we discuss spin and discrete symmetry constraints on operators
involving collinear fields.

The possible spin structures of currents with $\xi_{n,p}$ fields is restricted
by the fact that they have only two components, $\nslash \xi_{n,p}=0$.  The four
most general spin structures for currents with two collinear particles
moving in the same or opposite directions are
\begin{eqnarray} \label{Jspin}
  && \bar\xi_{n,p'}\,\Gamma_1\, \, \xi_{n,p}\qquad\quad 
  \Gamma_1 = \big\{ \bnslash\,, \bnslash\gamma_5\,,
                  \bnslash\gamma^\mu_\perp \big\}\,, \nn\\
  && \bar\xi_{\bn,p'} \,\Gamma_2\, \, \xi_{n,p}\qquad\quad 
  \Gamma_2 = \big\{ 1 \,, \gamma_5 \,, \gamma^\mu_\perp \big\}\,.
\end{eqnarray}
Other choices for $\Gamma_{1}$ and $\Gamma_2$ either vanish between the fields
or are related to those in Eq.~(\ref{Jspin}). This result can be expressed in a 
compact way by the trace formulae
\begin{eqnarray} \label{tracefomulae}
 && \bar\xi_{n,p'}\,\Gamma\, \xi_{n,p}\,,\qquad\quad 
 \Gamma = \frac{\bnslash}{8} {\rm Tr}[\nslash \Gamma]
    -\frac{\bnslash\gamma_5}{8} {\rm Tr}[\nslash\gamma_5 \Gamma] 
    -\frac{\bnslash\gamma_\perp^\mu}{8} {\rm Tr}[\nslash\gamma_\mu^\perp 
     \Gamma] \,,\nn \\
 && \bar\xi_{\bn,p'} \,\Gamma \, \xi_{n,p}\,,\qquad\quad
 \Gamma = \frac{1}{8} {\rm Tr}[\nslash\bnslash \Gamma]
    +\frac{\gamma_5}{8} {\rm Tr}[\gamma_5\nslash\bnslash \Gamma] 
    +\frac{\gamma_\perp^\mu}{8} {\rm Tr}[\gamma_\mu^\perp\nslash\bnslash 
     \Gamma] \,,
\end{eqnarray}
which reduce a general $\Gamma$ to a linear combination of the terms in
Eq.~(\ref{Jspin}).  For instance, it implies that $2i\bar\xi_{n} \sigma^{\mu\nu}
\xi_n =n^\nu \bar\xi_{n} \bnslash\gamma_\perp^\mu\xi_n -n^\mu \bar\xi_{n}
\bnslash \gamma_\perp^\nu \xi_n$, and $\bar\xi_{\bn} \gamma_\perp^\mu
\gamma_5\xi_n= i\epsilon_\perp^{\mu\nu} \bar\xi_{\bn} \gamma^\perp_\nu \xi_n$
where $\epsilon_\perp^{\mu\nu}=\epsilon^{\mu\nu\alpha\beta} \bn_\alpha
n_\beta/2$. Furthermore, each of the two components of $\xi_n$ and also
$\xi_{\bn}$ can be chosen to be eigenstates of their helicity operators,
$h={\hat p}\cdot{\vec S}$ with eigenvalues $\pm 1/2$.  For these fields $h$ is
equivalent to the chiral rotation, $h=\gamma_5/2$. The structures in
Eq.~(\ref{Jspin}) split into two classes depending on whether they conserve or
flip the helicity
\begin{eqnarray}
 \mbox{chiral even:}&& \mbox{\hspace{0.8cm}} 
  \bar\xi_{n,p'} \{\bnslash,\bnslash\gamma_5\} \xi_{n,p}\,,\qquad\, 
  \bar\xi_{\bn,p'} \gamma_\perp^\mu \xi_{n,p} \,, \\
  \mbox{chiral odd:}&& \mbox{\hspace{0.8cm}} 
  \bar\xi_{n,p'} \bnslash  \gamma_\perp^\mu\xi_{n,p}\,, \qquad\qquad
  \bar\xi_{\bn,p'} \{1,\gamma_5\} \xi_{n,p}\,.\nn
\end{eqnarray}
Since gluon interactions in QCD preserve helicity, integrating out hard QCD
fluctuations results in effective theory operators with the same helicity
structure as the original operators at leading order in $\lambda$.

The presence of labels on the effective theory fields makes their transformation
properties under the discrete symmetries $C$, $P$, and $T$ slightly different
than in QCD. For example under $P$ or $T$ we have $n\leftrightarrow \bn$ so
these transformations relate collinear fields for different directions.  Under
charge conjugation, parity, and time-reversal the collinear fields transform as
\begin{eqnarray}\label{CPT}
 \begin{array}{llll}
 & C^{-1} \xi_{n,p}(x) C = - \big[ \bar \xi_{n,-p}(x)\,{\cal C} \big]^T\,,
 && C^{-1} A^\mu_{n,p}(x) C = - [A_{n,p}^\mu(x)]^T \,, \\[5pt]
 & P^{-1} \xi_{n,p}(x) P =  \gamma_0\, \xi_{\bn,\tilde p}(x_P)\,, 
 &&  P^{-1} A^\mu_{n,p}(x) P =  g_{\mu\nu}A_{\bn,\tilde p}^\nu(x_P) \,,\\[5pt]
 & T^{-1} \xi_{n,p}(x) T =  {\cal T} \xi_{\bn,\tilde p}(x_T)\,, 
 &&  T^{-1} A^\mu_{n,p}(x) T =  g_{\mu\nu} A_{\bn,\tilde p}^\nu(x_T) \,,
 \end{array}
\end{eqnarray}
where ${\cal C}^{-1}\gamma_\mu{\cal C}=-\gamma_\mu^T$ and ${\cal
T}=\gamma^5{\cal C}$, while if $x^\mu=(x^+,x^-,x^\perp)$ and
$p^\mu=(p^+,p^-,p^\perp)$ then $\tilde p^\mu\equiv (p^-,p^+,-p_\perp)$,
$x_P^\mu\equiv (x^-,x^+,-\vec x^\perp)$, and $x_T^\mu\equiv (-x^-,-x^+,\vec
x^\perp)$. The transformation properties of $W_n$ can be worked out using
Eq.~(\ref{CPT}), for instance $C^{-1} W_n C = [W_n^\dagger]^T$.

The collinear effective Lagrangian (\ref{Lc}) is invariant under the
transformations in Eq.~(\ref{CPT}) (adding the $\bn\leftrightarrow n$
terms). These symmetries also constrain the form of non-perturbative matrix
elements. As an example, for a meson which is an eigenstate of $C$ one finds
\begin{eqnarray} \label{PSunderC}
 && \big\langle M_{n}\big| \bar{\xi}_{n,p} W_n \bnslash\gamma_5
  \delta(\omega-\bnP_+ ) W_n^\dagger \xi_{n,p'}\big| 0 \big\rangle \nn\\
 &&\qquad\qquad\quad 
  = (-1)^C \big\langle M_{n}\big| ({\cal C}W_n^\dagger\xi_{n,p})^T 
  \bnslash\gamma_5 \delta(\omega-\bnP_+) (\bar\xi_{n,p'} W_n{\cal C})^T 
  \big| 0 \big\rangle\nonumber\\
 &&\qquad\qquad\quad
  = (-1)^C \big\langle M_{n}\big| \bar{\xi}_{n,p'} W_n 
  \bnslash\gamma_5 \delta(\omega+ \bnP_+ ) W_n^\dagger \xi_{n,p} \big| 0
  \big\rangle\,.
\end{eqnarray}
For the isotriplet pion state $(-1)^C=+1$ so combining Eq.~(\ref{PSunderC}) with
Eq.~(\ref{pimom}) gives
\begin{eqnarray} \label{piC}
 && \big\langle \pi^a_{n,p}\big| \bar{\xi}_{n,p_1} W \Gamma_\pi^b
  \delta(\omega-\bnP_+ ) W^\dagger \xi_{n,p_2} \big| 0 
  \big\rangle \nn\\
 &&\qquad\qquad 
  = -if_\pi \delta^{ab}\, \bn\mcdot p  \int_0^1\!\! \mbox{d}x\, 
  \delta[-\omega \!-\! (2x\!-\!1)\bn\mcdot p ] \phi_\pi(x)\nn \\
 &&\qquad\qquad   
  =-if_\pi\delta^{ab}\, \bn\mcdot p  \int_0^1\!\! \mbox{d}x\, 
  \delta[\omega \!-\! (2x\!-\!1)\bn\mcdot p ] \phi_\pi(1\!-\!x) \,.
\end{eqnarray}
Together with Eq.~(\ref{pimom}), charge conjugation therefore implies that
$\phi_\pi(1-x)=\phi_\pi(x)$.

\section{Exclusive Processes} \label{section_exclusive}

\subsection{$\pi$-$\gamma$ Form Factor} \label{section_pionphoton}

The pion-photon form factor $F_{\pi\gamma}(Q^2)$ is perhaps the simplest setting
for factorization since there is only one hadron in the external state.  The
form factor is measurable in single-tagged two photon $e^-e^-\rightarrow e^-e^-
\pi^0$ reactions.  This process involves the scattering of a highly virtual
photon and a quark-anti-quark constituent pair off an on-shell photon.  The
photon scatters the quark pair away from the incoming photon into a pion, so
that $\gamma^*\gamma\to \pi^0$. The matrix element for this transition defines
the $\pi$-$\gamma$ form factor
\begin{eqnarray} \label{Fdef}
   \big\langle \pi^0(p_\pi) \big| J_\mu(0) \big| \gamma(p_\gamma,\epsilon)
      \big\rangle
  &=& ie\, \epsilon^\nu  \int\!\! d^4z\, e^{-ip_\gamma\cdot z} \big\langle
    \pi^0(p_\pi) \big|  {\rm T} J_\mu(0) J_\nu(z) \big| 0 
\big\rangle\nn \\
  &=&-ie\, F_{\pi\gamma}(Q^2)\epsilon_{\mu \nu \rho \sigma}p_\pi^\nu
    \epsilon^\rho\, q^\sigma \,.
\end{eqnarray}
Here $J_\mu= \bar\psi \cal{\hat Q} \gamma_\mu\psi$ is the full theory
electromagnetic current with isodoublet field $\psi$ and charge matrix $\hat
{\cal Q}= {\tau_3}/2+{\bf 1}/6$, and $-q^2=Q^2\gg \Lambda_{\rm QCD}^2$ where
$q=p_\pi-p_\gamma$ is the virtual photon momentum.  It has been shown that the
form factor can be written as a one-dimensional convolution of a hard
coefficient with the light-cone pion wavefunction~\cite{Pink}. Here we show how
this factorization takes place in the SCET.

\begin{figure}[t!]
   \centerline{ \hspace{-0.5cm}
   \includegraphics[width=1.5in]{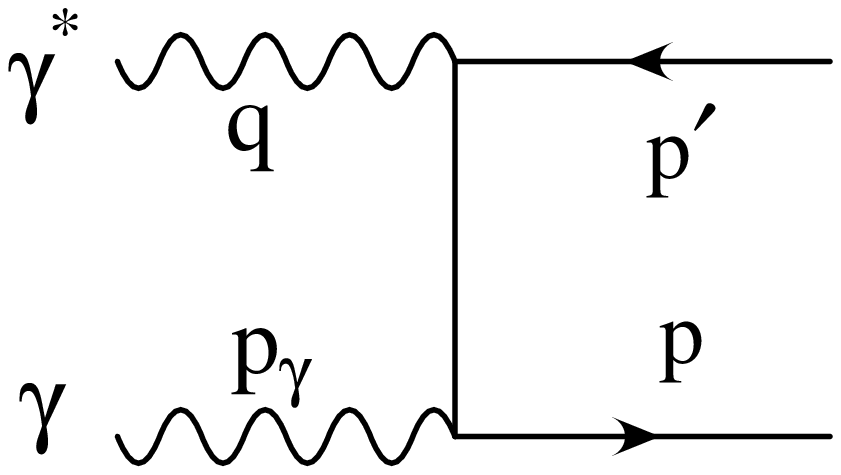}
   \quad \raisebox{1.3cm}{+}\quad
   \includegraphics[width=1.5in]{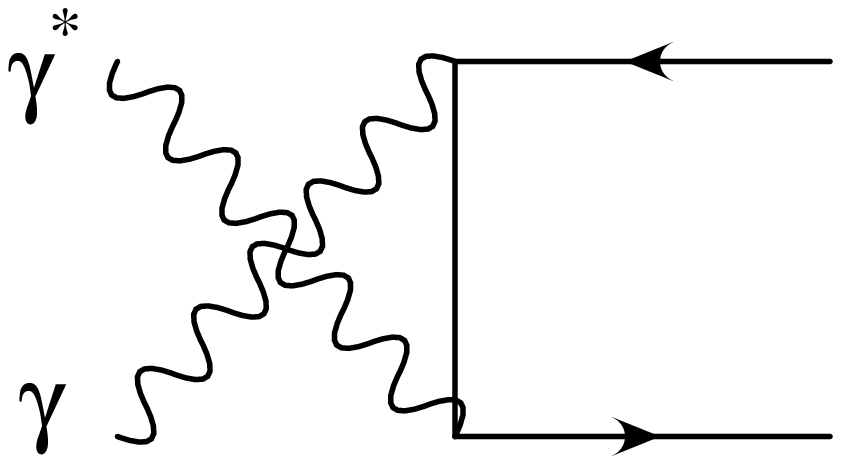}
   \quad\raisebox{1.3cm}{$\Longrightarrow$}\quad
   \includegraphics[width=1.5in]{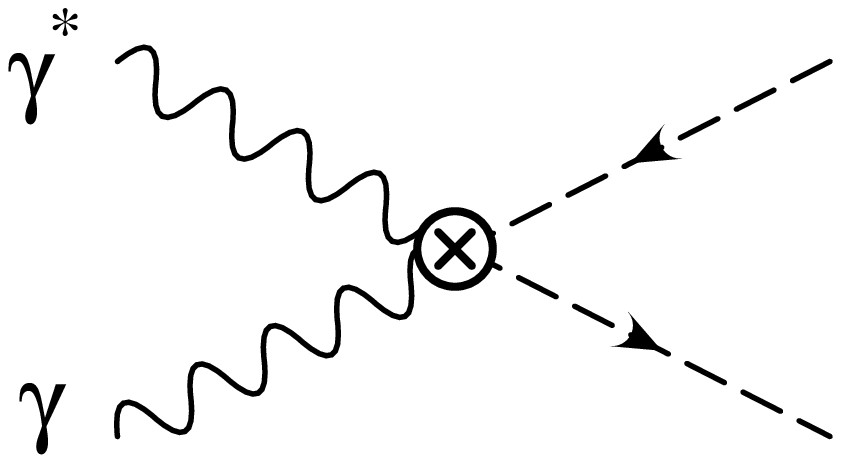}}
  \vspace{-0.4cm}
{\caption{Tree level matching onto $O_{j}$ in the Breit Frame. 
The graphs on the left include $u$ and $d$ quarks. \label{fig_gamgampi}}}
\end{figure}

In the Breit frame $q^\mu=Q(n^\mu-\bn^\mu)/2$, the real photon's momentum is
$p_\gamma^\mu=E \bn^\mu \simeq Q \bn^\mu/2$, and the pion is made up of
collinear particles with momenta $\bn\mcdot p_i \simeq Q$. The particles
exchanged between the two currents in Eq.~(\ref{Fdef}) have hard momenta and can
be integrated out. At leading order in $\lambda$ the time ordered product of the
two currents in Eq.~(\ref{Fdef}) matches onto a single operator in the
effective theory. For simplicity we restrict ourselves to the tensor and spin
structures that are relevant when the meson is a pion~\footnote{Note that a 
pure glue operator would not have the same isospin as the pion state.}
\begin{eqnarray} \label{Opig}
  O &=& \frac{i}{Q}\,\epsilon_{\mu\nu}^\perp\ [\bar{\xi}_{n,p}W]\,\Gamma
   C(\bnP,\bnPd, \mu)\, [W^\dagger \xi_{n,p^\prime}]  \,,
\end{eqnarray}
where $\xi_{n,p}$ is an isodoublet collinear quark field, and
$2\epsilon_{\mu\nu}^\perp = \epsilon_{\mu\nu\rho\beta} \bn^\rho n^\beta$. $
O_{1}$ is of dimension two, just like the time-ordered product in
Eq.~(\ref{Fdef}), and a power of $1/Q$ is included to make $C(\mu,\bnP,\bnPd)$
dimensionless. The time-ordered product in Eq.~(\ref{Fdef}) is even under charge
conjugation, so the operators in Eq.~(\ref{Opig}) must also be even.  This
implies $C_{\pi\gamma}(\mu,\bnP,\bnPd) = C_{\pi\gamma}(\mu,-\bnPd,-\bnP)$. The
location of the $W$'s in Eq.~(\ref{Opig}) is fixed by gauge invariance, and
$\Gamma$ contains the spin and flavor structure
\begin{eqnarray} \label{Gpig}
  \Gamma &=& \big( {\bnslash}\gamma_5\big) \big(3\sqrt{2}\,\hat {\cal Q}^2\big)
  \,.
\end{eqnarray}

Since the offshellness of the collinear particles in the pion is $p^2\sim
\Lambda_{\rm QCD}^2$ we can also integrate out offshell modes with $p^2\sim
Q\Lambda_{\rm QCD}$ which come from soft-collinear interactions.  For the
collinear operators $O_j$, Eq.~(\ref{Sadd}) implies that factors of the soft
Wilson line $S_n$ are induced. However, the location is such that $S^\dagger_n
S_n=1$, so no coupling to soft gluons occur at leading order. The coupling of
the collinear fields to usoft gluons can be simplified with the field
redefinitions in Eq.~(\ref{def0}). As discussed in section~\ref{section_SCET}
this moves all couplings into $O_j$, and using $Y_n^\dagger Y_n=1$ gives
\begin{eqnarray}
  O &=& \frac{i}{Q}\,\epsilon_{\mu\nu}^\perp\ [\bar{\xi}_{n,p}^{(0)}W^{(0)}]
   \,\Gamma\, C(\bnP,\bnPd, \mu)\, [W^{(0)\dagger} \xi_{n,p^\prime}^{(0)}]
   \,.
\end{eqnarray}
Thus, usoft gluons also decouple.

In the Breit frame the pion momentum satisfies $p^\mu_{\pi}= E_\pi n^\mu+{\cal
O}(\lambda)$, and comparing Eq.~(\ref{Fdef}) with the SCET matrix element
$i\langle \pi^0_{n,p_\pi} | O_{\pi\gamma} |0\rangle$, gives
\begin{eqnarray} \label{pig2}
\frac{Q^2}{2} F_{\pi\gamma}(Q^2)
  &=& \frac{i}{Q}\, \big\langle \pi^0_{n}\big|\bar{\xi}^{(0)}_{n,p}
   W^{(0)}\,\Gamma\, C(\bnP,\bnPd,\mu)\, W^{(0)\dagger} 
  \xi^{(0)}_{n,p^\prime} \big| 0 \big\rangle \,.
\end{eqnarray}
Defining $\bnP_{\pm}=\bnPd \pm \bar{\cal P}$, the operator $\bnP_-$ is related
to $\bnP$ acting from the outside on the fields. Using Eq.~(\ref{MC}) it can
therefore be set equal to the momentum label of the state, $\bnP_- =\bn\cdot
p_\pi = Q$. Suppressing this dependence we write
$C(\bnP,\bnPd,\mu)=C_1(\bnP_+,\mu)$ leaving
\begin{eqnarray} \label{pig3}
  F_{\pi\gamma}(Q^2) &=& \frac{2i}{Q^3} \big\langle \pi^0_{n} \big| \,
   \bar{\xi}_{n,p}^{(0)} W^{(0)} \Gamma \,  C_1(\bnP_+,\mu) W^{(0)\dagger}
   \xi_{n,p\prime}^{(0)} \, \big| 0 \big\rangle \nn \\[3pt]
   &=& \frac{2i}{Q^3} \int \!\!{d\omega}\,C_1(\omega,\mu)
   \big\langle \pi^0_{n} \big|  \bar{\xi}_{n,p}^{(0)} W^{(0)}
   \Gamma \delta(\omega-\bnP_+) W^{(0)\dagger} \xi^{(0)}_{n,p\prime}
   \big| 0 \big\rangle \,.
\end{eqnarray}
Using Eq.~(\ref{pimom}) the remaining matrix identity in Eq.~(\ref{pig3}) can be
written in terms of the light-cone pion wavefunction
\begin{eqnarray} \label{resultpionformfactor}
  F_{\pi\gamma}(Q^2) &=& \frac{2 f_\pi}{Q^2} \int\!\! d\omega\!
  \int_0^1\!\! dx\ \delta(\omega-(2x-1)2E_\pi)\:
  C_1(\omega,\mu)\: \phi_{\pi}(x,\mu)   \nn \\
  &=& \frac{2f_\pi}{Q^2} \int_0^1\!\! dx\: C_1((2x-1)Q,\mu)
  \: \phi_{\pi}(x,\mu) \,.
\end{eqnarray}
This is the final result and is valid to leading order in $\lambda$ and all
orders in $\alpha_s$.  From Eq.~(\ref{piC}) charge conjugation implies that
$\phi_\pi(x) = \phi_\pi(1-x)$ and $C_1(\omega) = C_1(-\omega)$.
Eq.~(\ref{resultpionformfactor}) agrees with the
Brodsky-Lepage~\cite{brodskylepage} result that the form factor can be written
as the convolution of a short distance function with the light-cone pion
wavefunction.  The SCET formalism gives a concise derivation of this result and
defines the short distance function in terms of the Wilson coefficient of an
effective theory operator.

As an illustrative example consider the tree level matching onto $C$
illustrated in Fig.~\ref{fig_gamgampi}. Since the location of the $W$'s in $O$
are fixed by gauge invariance, $C(\mu,\bnP,\bnPd)$ can be determined by
matching with $W=1$.  Expanding the full theory graphs to leading order gives
\begin{eqnarray} \label{1ab}
i\: (\mbox{Fig.}1) =
   \frac{ie}{2}\,\epsilon_{\mu\nu\rho\beta}\epsilon^\nu \bn^\rho n^\beta
   \Big(\frac{\bnslash}{2}\gamma_5\Big) \big({\cal{\hat Q}}^2\big)
   \left( \frac{1}{\bn\cdot p}-\frac{1}{\bn \cdot p^\prime} \right),
\end{eqnarray}
where we have dropped isosinglet terms, contributions with opposite parity to
the pion, as well as those proportional to $\nslash\gamma_5$ since $\nslash
\xi_{n,p}=0$.  Comparing Eq.~(\ref{1ab}) to Eq.~(\ref{Opig}) gives
\begin{eqnarray} \label{Cpig_tree}
   C(\mu,\bnP,\bnPd)
    =\frac{Q}{6\sqrt{2}} \Big( \frac{1}
  {\bnPd}-\frac{1}{\bnP}\Big) + {\cal O}\big(\alpha_s(Q)\big)  \,,
\end{eqnarray}
so that
\begin{eqnarray} \label{Cpig2}
   C_1(\mu,\omega=(2x-1)Q)
    =\frac{1}{6\sqrt{2}} \Big( \frac{1}{x}+\frac{1}{1-x}\Big)
     + {\cal O}\big(\alpha_s(Q)\big)  \,.
\end{eqnarray}
This result is again in agreement with Ref.~\cite{brodskylepage}, and the order
$\alpha_s(Q)$ corrections to this Wilson coefficient can be read off from the
results in Ref.~\cite{delAguila:1981nk,Braaten:yp}. An identical analysis
applies for operators with different spin structures such as the ones
contributing to $\gamma^*\gamma\to\rho^0$.

\subsection{The large $Q^2$ meson form factor}

Another example of an exclusive process which can be treated in the effective
theory is the classic case of the electromagnetic pion form factor at large
space-like momentum transfer. For generality we consider in this section the
electromagnetic form factors for arbitrary mesons (pseudoscalar $P$ or vector
$V$), defined as
\begin{eqnarray}\label{piff}
& &\big\langle P^\prime(p^\prime)\big| J_\mu \big| P(p) \big\rangle
  = F_P(Q^2) (p_\mu+p^\prime_\mu)\,,\\
& &\big\langle V^\prime(p^\prime,\varepsilon^\prime) \big| J_\mu
  \big| P(p) \big\rangle
  =  G(Q^2) i\epsilon_{\mu\nu\alpha\beta}\, p^\nu p^{\prime\alpha}
    \varepsilon^{\prime\beta} \nn\,, \\
& &\big\langle V^\prime(p^\prime,\varepsilon^\prime) \big|J_\mu
   \big| V(p,\varepsilon)\big\rangle
   = F_1(Q^2) (\varepsilon^{\prime*}\cdot\varepsilon)(p+p^\prime)_\mu +
     F_2(Q^2) [(\varepsilon^{\prime*}\cdot p)\varepsilon_\mu +
     (\varepsilon \cdot p^\prime)\varepsilon^{\prime*}_\mu]\,,\nn
\end{eqnarray}
where $q^2 = -Q^2, q=p-p^\prime$. For simplicity we suppress the dependence of
the form factors on the isospin of the two mesons. We will restrict ourselves in
the following to the case of hadrons made up only of $u,d$ quarks.  The
electromagnetic current is defined as usual by $ J_\mu = \bar q\:\hat {\cal
Q}\,\gamma_\mu\: q$, with charge matrix $\hat {\cal Q} = \mbox{diag}(2/3,-1/3)$,
which can be written in terms of the up and down quark charges as $\hat {\cal
Q}=(Q_u-Q_d)\tau_3/2+(Q_u+Q_d){\bf 1}/2$.

We will be interested in the asymptotic form of the form factors in the region
with $Q^2 \gg m_P^2$, where it can be expanded in a power series in $1/Q^2$
\cite{brodskylepage}.  It is convenient to work in the Breit frame, where the
momentum transfer has the light-cone components $q=(q^+, q^-, q_\perp) = (Q, -Q,
0)$. In this frame the meson momenta are $p = (Q, {m_P^2}/{Q}, 0)$, $p'=
({m_P^2}/{Q}, Q, 0)$, so the partons in the incoming/outgoing meson are
collinear along the $\bn_\mu$/$n_\mu$ direction.

\begin{figure}[t!]
   \centerline{ \hspace{-0.5cm}
   \includegraphics[width=1.5in]{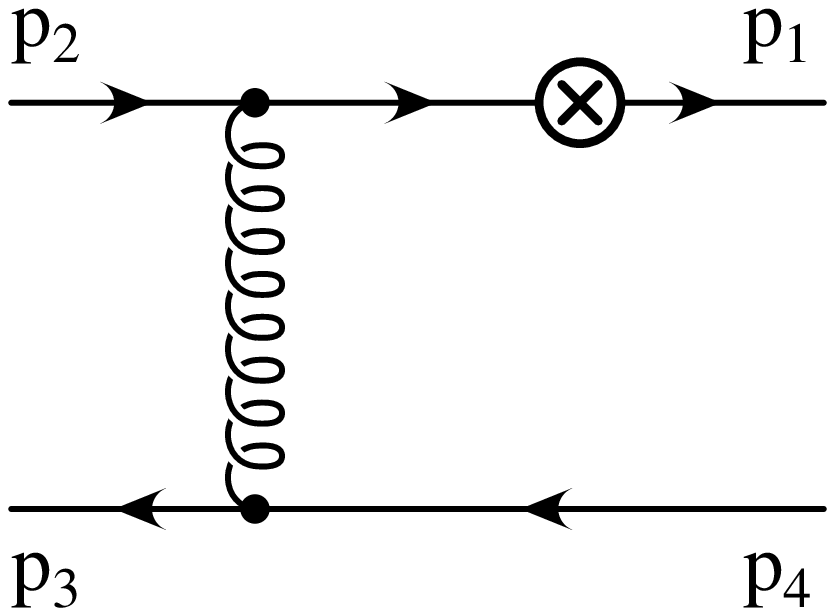}
   \hspace{0.3cm}\raisebox{1.3cm}{+}\hspace{0.4cm}
   \includegraphics[width=1.5in]{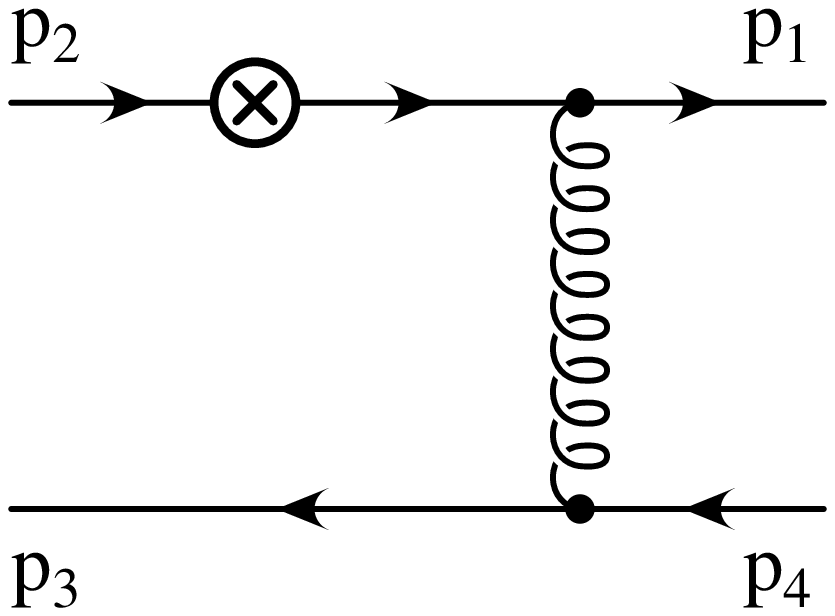}
   \quad\raisebox{1.3cm}{$\Longrightarrow$}\hspace{0.4cm}
   \includegraphics[width=1.5in]{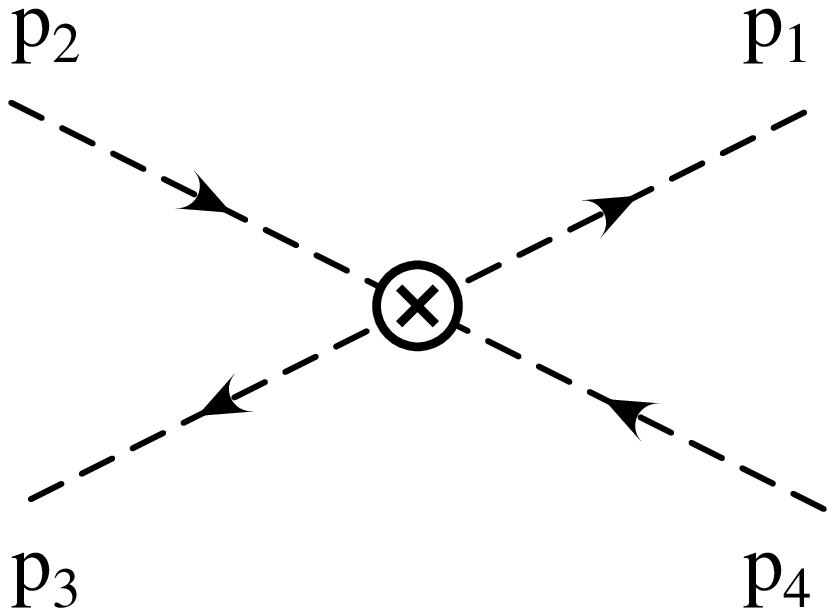}}
  \vspace{-0.4cm}
{\caption{Tree level matching onto $C_{0,8}$. The QCD graphs on the left 
plus the analogous graphs with the current on the bottom quark line
are matched onto the collinear operator on the right. \label{fig_pipi}}}
\end{figure}

The electromagnetic current in Eq.~(\ref{piff}) is matched in the effective
theory onto the most general combination of operators constructed from collinear
fields which are compatible with collinear gauge invariance.  Operators such as
the dimension three current
\begin{eqnarray} \label{Feyn}
    [\bar\xi_{\bn}W_{\bn}] \Gamma\: C(\mu,\cP^\dagger,\bnP)\: 
[W^\dagger_n \xi_n] \,,
\end{eqnarray}
can contribute, but only overlap with the asymmetric meson states with one
energetic collinear quark and one usoft or soft quark. Often this overlap is
referred to as the tail of the wavefunction contribution or the Feynman
mechanism of generating the form
factor~\cite{brodskyfeynman,isgurfeynman}. There are other operators with
significant overlap with more symmetric meson states (where all the constituents
are allowed to be energetic).  The leading such operators  have the
form~\footnote{There are also gluon operators that can contribute when one or 
more of the mesons is a neutral isosinglet, however for simplicity these are
not discussed here.} 
\begin{eqnarray} \label{BLmech}
   \frac{1}{Q^3}\,
   \big[\bar \xi_{n,p_1} W_n  \Gamma W^\dagger_{\bn} \xi_{\bn,p_2} \big]
   C(\mu,\bnP,\bnPd,\cP,\cP^\dagger)
   \big[ \bar \xi_{\bn,p_3} W_{\bn} \Gamma' W^\dagger_n \xi_{n,p_4} 
\big]\,,
\end{eqnarray}
with $C$ a dimensionless Wilson coefficient. As usual, collinear gauge
invariance is enforced by the location of the $W$'s in Eqs.~(\ref{Feyn}) and
(\ref{BLmech}).  There is some argument about the relative size of
Eqs.~(\ref{Feyn}) and (\ref{BLmech}) in the
literature~\cite{brodskyfeynman,isgurfeynman}. Often it is argued that the tail
of the wavefunction is suppressed by an extra $\Lambda_{\rm
QCD}^2/Q^2$~\cite{brodskyfeynman}, in which case the operator in
Eq.~(\ref{BLmech}) dominates by two powers of $Q$. An analysis of the tail of
wavefunction contributions has not yet been performed in the effective theory
framework. Therefore, we choose to ignore the operator in Eq.~(\ref{Feyn}), and
below only analyze the operator in Eq.~(\ref{BLmech}). We emphasize that we do
not claim to have shown that this is justified by the effective theory power
counting.

There are two different structures possible for the operator in
Eq.~(\ref{BLmech}), and we write the general matching for the electromagnetic
current as
\begin{eqnarray}\label{Jmatch}
  J^\nu \to  \frac{1}{Q^3}\,\int\!\!{d\omega_j} \:\Big[
    C_0(\mu,\omega_j)\: {\cal J}^\nu_0(\omega_j,\mu) +
    C_8(\mu,\omega_j)\: {\cal J}^\nu_8(\mu,\omega_j) \Big] \,,
\end{eqnarray}
where $j=1,2,3,4$.  The SCET currents are dimension-6 operators
\begin{eqnarray} \label{J12}
  {\cal J}_0^\nu &=& \bCH{n}{\omega_1} \Gamma \CH{\bn}{\omega_2}\:
   \bCH{\bn}{\omega_3} \Gamma' \CH{n}{\omega_4}
    -(\Gamma\leftrightarrow\Gamma',
     \omega_{1,2}\leftrightarrow-\omega_{4,3}) \,,\\
  {\cal J}_8^\nu &=&  \bCH{n}{\omega_1} \Gamma\, T^a \CH{\bn}{\omega_2}\:
   \bCH{\bn}{\omega_3} \Gamma' T^a \CH{n}{\omega_4}
    - (\Gamma\leftrightarrow\Gamma',
     \omega_{1,2}\leftrightarrow-\omega_{4,3}) \nn\,,
\end{eqnarray}
where the $\chi$ fields are defined in Eq.~(\ref{CH1}). In terms of the charge
matrix $\hat {\cal Q}$, the spin and flavor structure is
\begin{eqnarray}
  \Gamma\otimes \Gamma' = (n^\nu + \bn^\nu) \Big(
    \gamma_\alpha^\perp \hat {\cal Q}  \otimes \gamma^\alpha_\perp  {\bf 1} 
\Big) \,.
\end{eqnarray}
The Wilson coefficients $C_{0,8}$ can be computed in a power series in
$\alpha_s(Q)$. They are functions of $\mu$, $Q$, and the $\omega_j$ which are
the sum of momentum labels for gauge invariant products of collinear fields in
the SCET currents.

The currents operators in Eq.~(\ref{J12}) are the most general allowed operators
which are gauge invariant, transform the same way as $J^\mu$ under charge
conjugation and satisfy current and helicity conservation.  To see how these
properties constrain the form of the allowed operators, we begin by noting that
Eq.~(\ref{Jspin}) implies that $\Gamma,\Gamma'=\{1,\gamma_5,\gamma_\perp^\mu\}$
are the most general allowed spin structures. For massless quarks the
electromagnetic and QCD couplings preserve helicity, whereas $\bar \xi_n
\{1,\gamma_5\} \xi_{\bn}$ cause the helicity to flip. Thus, only the 
structure $\bar \xi_n \gamma_\perp^\mu \xi_{\bn}$ is allowed. Current
conservation $q^\nu {\cal J}_\nu=0$, together with $q^\nu = Q(\bn^\nu-n^\nu)/2$
implies ${\cal J}_\nu\propto (n_\nu+\bn_\nu)$. Under charge conjugation
$J^\mu\to -J^\mu$ so the same must be true for the SCET currents. In the current
operators, charge conjugation switches $\omega_1\leftrightarrow -\omega_4$,
$\omega_2 \leftrightarrow -\omega_3$, and $\Gamma \leftrightarrow \Gamma'$, as
can be seen from Eq.~(\ref{CPT}). Thus, the second term in ${\cal J}_{0,8}^\nu$
is required to make these operators odd under charge conjugation. The operators
${\cal J}_{0,8}^\nu$ and the full electromagnetic current are invariant under a
combined $PT$ transformation. This requires that the Wilson coefficients are
real.

The operators ${\cal J}_{0,8}$ are responsible for the $P_\bn\to P_n$
transition, while the reverse transition $P_n\to P_\bn$ is described by similar
operators with $\bn\leftrightarrow n$. Parity invariance requires the Wilson
coefficients of these operators to be identical to $C_{0,8}(\omega_i)$.
Demanding hermiticity of the electromagnetic current in the effective theory
then gives the relation $C_{0,8}(\omega_1, \omega_2, \omega_3, \omega_4)=
C_{0,8}^*(\omega_2, \omega_1, \omega_4, \omega_3)$.  Since the coefficients are
real they must therefore satisfy $C_{0,8}(\omega_1, \omega_2, \omega_3,
\omega_4)= C_{0,8}(\omega_2, \omega_1, \omega_4, \omega_3)$.

To compute the matrix elements in the effective theory, it is convenient to
Fierz transform the four-quark operators in Eq.~(\ref{J12}). This gives
\begin{eqnarray} \label{J12f}
   C_0 {\cal J}_0 + C_8 {\cal J}_8 = (n^\nu + \bn^\nu)\:
    \sum_{j=1}^4 C_j {\cal J}_j \,,
\end{eqnarray}
where
\begin{eqnarray} \label{JemC}
  {\cal J}_j &=& \big[\bCH{n}{\omega_1} \Gamma_j
   \CH{n}{\omega_4}\big] \big[ \bCH{\bn}{\omega_3} \Gamma_j^\prime 
   \CH{\bn}{\omega_2} \big]   \,.
\end{eqnarray}
The spin, flavor, and color structures are
\begin{eqnarray} \label{G12}
   \Gamma_1 \otimes \Gamma_1^\prime &=& -\frac14 (Q_u-Q_d) i\epsilon^{3bc}
   (\tau^b\otimes\tau^c)
   \big[ \bnslash\otimes \nslash + \bnslash\gamma_5\otimes 
\nslash\gamma_5 \big]
   \,,\nn\\
   \Gamma_2 \otimes \Gamma_2^\prime &=& \frac14 \Big[
   (Q_u+Q_d) ({\bf 1}\otimes {\bf 1} + \tau^a \otimes \tau^a) +
   (Q_u-Q_d) ({\bf 1}\otimes \tau^3 + \tau^3 \otimes {\bf 1}) \Big]\nn\\
   && \times   \big[ \bnslash\otimes \nslash
    + \bnslash\gamma_5\otimes \nslash\gamma^5 \big] \,,
\end{eqnarray}
while $\Gamma_{3,4}=T^a\Gamma_{1,2}$ and $\Gamma_{3,4}'=T^a\Gamma_{1,2}'$.  The
new Wilson coefficients are
\begin{eqnarray} \label{C12}
  C_1(\mu,\omega_j) &=& \bigg[ \frac18 \Big(1-\frac{1}{N_c^2}\Big)
   C_8(\mu,\omega_j) + \frac{1}{4N_c} C_0(\mu,\omega_j) \bigg]
   + (\omega_{1,2}\leftrightarrow -\omega_{4,3}) \,,\nn\\
  C_2(\mu,\omega_j) &=& \bigg[ \frac18 \Big(1-\frac{1}{N_c^2}\Big)
   C_8(\mu,\omega_j) + \frac{1}{4N_c} C_0(\mu,\omega_j) \bigg]
   - (\omega_{1,2}\leftrightarrow -\omega_{4,3}) \,,
\end{eqnarray}
with similar relations for $C_{3,4}$ which are also in terms of $C_{0,8}$.

A few general predictions follow from the form of the operators in
Eq.~(\ref{JemC}).\footnote{These predictions depend on the dominance of the
operators in Eq.~(\ref{BLmech}) over those in Eq.~(\ref{Feyn}).} For mesons with
spin, only helicity conserving form factors appear, and furthermore no
off-diagonal (e.g., $P\to V$) matrix elements are present at leading order in
$1/Q^2$. These results agree with Ref.~\cite{Pink}. We also see that the form
factors between arbitrary meson states are determined at leading power by only
two hard coefficients, $C_0$ and $C_8$.

Now consider what factorization tells us about the matrix element of the
operators in Eq.~(\ref{JemC}). For the decoupling of usoft and soft gluons we
will follow section~\ref{section_SCET}. Integrating out offshell modes with
$p^2\sim Q\Lambda_{\rm QCD}$ induces soft Wilson lines $S_n$ and $S_\bn$, while
the field redefinitions in Eq.~(\ref{def0}) make all couplings to usoft gluons
explicit in the operators. Together these give
\begin{eqnarray}
    {\cal J}_j^\nu &=& \big[ \bCH{n}{\omega_1}^{(0)} Y_n^\dagger S_n^\dagger
   \Gamma_j S_n Y_n  \CH{n}{\omega_4}^{(0)}\big]\:
   \big[ \bCH{\bn}{\omega_3}^{(0)} Y_\bn^\dagger S_\bn^\dagger \Gamma_j^\prime 
   S_\bn Y_\bn \CH{\bn}{\omega_2}^{(0)}\big]    \,.
\end{eqnarray}
Consider first the color singlet currents $j=1,2$. Here the $Y$'s and $S$'s all
cancel using unitarity of the Wilson lines.  Since the $A_{n,q}^\mu$ and
$A_{\bn,q}^\mu$ gluons only interact with fields in the $n$ and $\bn$ directions
respectively, collinear gluons are not exchanged between the $n$ and $\bn$ quark
bilinears.  Thus, the matrix element between states with particles moving in the
$n$ and $\bn$ directions factors
\begin{eqnarray}\label{Jfactor}
   \langle n | {\cal J}_{1,2} | \bn \rangle =
   \langle n | \bCH{n}{\omega_1}^{(0)} \Gamma_{1,2} \CH{n}{\omega_4}^{(0)}
   | 0\rangle
   \langle 0 | \bCH{\bn}{\omega_3}^{(0)} \Gamma_{1,2}^\prime
   \CH{\bn}{\omega_2}^{(0)}| \bn \rangle \,.
\end{eqnarray}
Next consider the currents ${\cal J}_{3,4}$, which have color structure $T^a
\otimes T^a$ in $\Gamma_j\otimes \Gamma_j^\prime$. In this case the usoft and
soft gluons do not cancel, but can all be moved into one quark bilinear using
the color identity $Y_n^\dagger S_n^\dagger T^a S_n Y_n \otimes Y_\bn^\dagger
S_\bn^\dagger T^a S_\bn Y_\bn = T^a \otimes Y_\bn^\dagger S_\bn^\dagger S_n Y_n
T^a Y^\dagger_n S^\dagger_n S_\bn Y_\bn$. After this rearrangement it is clear
that the (u)soft gluons and $A_{n,q}^\mu$ and $A_{\bn,q}^\mu$ gluons only
interact with the fields in one of the quark bilinears. Thus, the matrix element
$\langle n |{\cal J}_{3,4}|\bn \rangle$ factors, similar to
Eq.~(\ref{Jfactor}). For color singlet states, however, the matrix element of an
octet operator vanishes identically since
\begin{eqnarray}
   \langle n | \bCH{n}{\omega_1}^{(0)} T^a \CH{n}{\omega_4}^{(0)}
   | 0\rangle =0 \,.
\end{eqnarray}
Thus, the effective theory currents ${\cal J}_{3,4}$ do not contribute to the
form factors at any order in perturbation theory.

Eq.~(\ref{Jfactor}) shows that for arbitrary meson states factorization occurs.
It remains to show that the matrix elements in Eq.~(\ref{Jfactor}) are given by
a two-dimensional convolution with the light-cone meson wavefunctions. To do
this we consider the simple example of the $0^{-+}\to 0^{-+}$ form factor for
the charged pion. It should be obvious that the same steps go through for other
meson states.  

The symmetry of the pion wavefunction $\phi_\pi(x)$ under charge conjugation
($x\to 1-x$) implies that only the ${\cal J}_1$ current contributes.  Thus,
\begin{eqnarray} 
   F_{\pi^\pm}(Q^2) =  \frac{2}{Q^4} \int\!\!{d\omega_j}\,
  C_1(\mu,\omega_j) \big\langle \pi_n^\pm(p') \big| \bCH{n}{\omega_1}^{(0)}
  \Gamma_1 \CH{n}{\omega_4}^{(0)} \big| 0 \big\rangle
  \big\langle 0 \big| \bCH{\bn}{\omega_3}^{(0)} \Gamma_1' 
  \CH{\bn}{\omega_2}^{(0)} \big| \pi_\bn^\pm(p) \big\rangle \,.
\end{eqnarray}
The required matrix elements can be obtained from Eqs.~(\ref{pimom}) and
(\ref{pimom2}) with $|\pi^\pm\rangle = \mp(\pi^1 \pm i\pi^2)/\sqrt{2}$. The
momentum conserving delta functions fix $\omega_1 -\omega_4 = \bn\cdot p'=Q$ and
$\omega_2-\omega_3=\bn\cdot p=Q$, while the $\omega=\omega_1+\omega_4$ and
$\omega'=\omega_3+\omega_2$ integrations can be done with the delta
functions. This leaves
\begin{eqnarray}
F_{\pi^\pm} &=& \pm(Q_u\!-\!Q_d)\, \frac{f_\pi^2}{Q^2}\, 
\int_0^1\!\!\mbox{d}x\!
  \int_0^1\!\!\mbox{d}y \: T_1(x,y,\mu) \phi_\pi(x,\mu) \phi_\pi(y,\mu)\,,
\end{eqnarray}
where $T_1(x,y)$ is defined in terms of
$C_1(\omega_1,\omega_2,\omega_3,\omega_4)$ as
\begin{eqnarray}
  T_1(x,y) = C_1\big(xQ,yQ,(y-1)Q,(x-1)Q)
  \,.
\end{eqnarray}

The coefficients $C_{j}(\mu,\omega_j)$, and therefore also $T_j(x,y)$, can be
obtained at the scale $\mu=Q$ by a matching calculation, as illustrated in
Fig.~\ref{fig_pipi}. For this purpose, it is sufficient to compute the matrix
element of the currents with free collinear quarks.  To lowest order in
$\alpha_s(Q)$, only $C_8(\omega_j,\mu=Q)$ is nonvanishing
\begin{eqnarray} \label{piff20}
   C_0(\omega_j,\mu=Q) =0\,,\qquad\quad
   C_8(\omega_j,\mu=Q)  = 4\pi \alpha_s(Q)\: \frac{Q^2}{\omega_3\, 
\omega_4 }\,.
\end{eqnarray}
This implies
\begin{eqnarray}
  T_1(x,y,\mu=Q) = \frac{4\pi \alpha_s(Q)}{9} \Big[
    \frac{1}{xy}+\frac{1}{(1-x)(1-y)} \Big] \,,
\end{eqnarray}
Using the asymptotic light-cone pion wavefunction
$\phi_\pi(x)=6x(1-x)$ we find agreement with Ref.~\cite{brodskylepage},
\begin{eqnarray} \label{piff2}
F_{\pi^\pm}(Q^2) = \pm \frac{8\pi f_\pi^2 \alpha_s(Q)}{9Q^2}
\left[ \int_0^1\!\!\mbox{d}x\, \frac{\phi_\pi(x)}{x} \right]^2 \to \pm
\frac{8\pi f_\pi^2 \alpha_s(Q)}{Q^2}\,.
\end{eqnarray}
The order $\alpha_s^2(Q)$ corrections to Eq.~(\ref{piff20}) can be found in
Refs.~\cite{Field:wx,Dittes:aw,Braaten:yy}.


\section{Inclusive Processes}\label{section_inclusive}

\subsection{Deep inelastic scattering}

DIS is a process which is both simple and rich in physics. As such it provides
an ideal introduction to inclusive factorization in QCD, which we study from an
effective field theory point of view in this section. The aim is to prove that
to all orders in $\alpha_s$ and leading order in $\lambda$ the DIS forward
scattering amplitude can be written as an integral over hard coefficients times
the parton distribution functions. This is done by matching onto local operators
in SCET. The properties of SCET are used to show that matrix elements of the
leading local operator can be written as a convolution of a hard coefficient
with the parton distribution functions for the proton.

The first step is to understand the kinematics of the process. The hard scale
$Q^2 =-q^2$ is set by the invariant mass of the photon, and $x = Q^2/(2 p\mcdot
q)$ is the Bjorken variable. In the Breit frame the momentum of the virtual
photon is $q^\mu=Q(\bn^\mu-n^\mu)/2$, and the incoming proton momentum is
$p^\mu= n^\mu \bn\mcdot p/2 + \bn^\mu m_p^2/(2\bn\mcdot p) \simeq n^\mu Q/(2x)
+\bn^\mu x m_p^2/(2Q)$ up to terms $\sim m_p^2/Q^2$, where $m_p$ is the proton
mass. By momentum conservation the final state momentum is $P^\mu_X =
q^\mu+p^\mu$, which gives an invariant mass $P_X^2 = (Q^2/x)
(1-x)+m_p^2$. Values $1-x\simeq \Lambda_{\rm QCD}/Q$ correspond to the endpoint
region where the particles in $X$ are collimated into a jet, while values
$1-x\simeq \Lambda_{\rm QCD}^2/Q^2$ correspond to the resonance region. We will
consider the standard OPE region where $1-x \gg \Lambda_{\rm QCD}/Q$ so that the
final state has virtuality of order $Q^2$ and can be integrated out. In
contrast, although the incoming proton has a large momentum component in the
$n^\mu$ direction it has a small invariant mass $p^2 = m_p^2 \sim \Lambda_{\rm
QCD}^2$, and therefore is described by collinear fields in the effective theory.

Consider the spin-averaged cross section for DIS which can be written as
\begin{eqnarray}
 d\sigma = \frac{d^3 \mathbf{k}'}{2|\mathbf{k}'|(2\pi)^3}\: 
 \frac{\pi e^4}{s Q^4}\: L^{\mu\nu}(k,k')\: W_{\mu\nu}(p,q) \,,
\end{eqnarray}
where $k$ and $k'$ are the incoming and outgoing lepton momenta with $q=k'-k$,
$L^{\mu\nu}$ is the lepton tensor, and $s=(p+k)^2$.  The hadronic tensor
$W^{\mu\nu}$ can be related to the imaginary part of the DIS forward scattering
amplitude:
\begin{eqnarray}
 W_{\mu\nu}(p,q) &=& \frac{1}{\pi}\, {\rm Im}\, T_{\mu\nu}(p,q) \,,\\
 T_{\mu\nu}(p,q) &=& \frac12 \sum_{\rm spin} \big\langle p \big| 
  \hat T_{\mu\nu}(q) \big| p \big\rangle\,,\qquad\quad 
  \hat T_{\mu\nu}(q)=i\int d^4z\: e^{iq\cdot z}\,{\rm T}[J_\mu(z)J_\nu(0)]\,,\nn
\end{eqnarray}
where for an electromagnetic current $J_\mu$ we can write
\begin{eqnarray} \label{tprod}
  T_{\mu\nu}(p,q) = \Big(-g_{\mu\nu} + \frac{q_\mu q_\nu}{q^2}\Big)
  T_1(x,Q^2)+\left(p_\mu+\frac{q_\mu}{2x}\right)
  \left(p_\nu+\frac{q_\nu}{2x}\right)T_2(x,Q^2) \,. 
\end{eqnarray}

As explained above, the intermediate hadronic state has invariant mass $P_X^2
\sim Q^2$. Therefore, one can perform an OPE and match $\hat T^{\mu\nu}(q)$ onto
operators in SCET. All fields in the resulting operators are evaluated at the
same residual space time point, however, the presence of Wilson lines and label
momenta make the operators nonlocal along a particular light cone
direction. These nonlocal operators sum the infinite set of purely local
operators of a given twist, however this is built into the formalism
automatically.
\begin{figure}[t!]
  \includegraphics[width=5.5in]{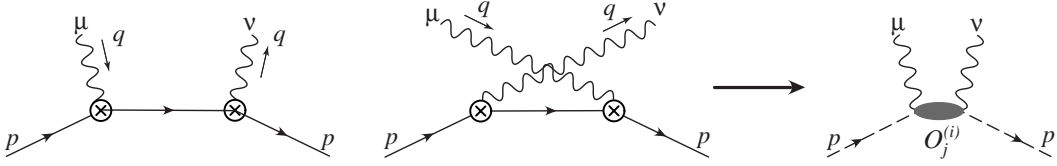}
{\caption{Tree level matching onto the operator $O_j^{(i)}$ in DIS.
\label{disopefig}}}
\end{figure}
To match we write down the most general leading operator in SCET which contains
collinear fields moving in the $n^\mu$ direction, and enforce the condition from
current conservation $q^\mu \hat T_{\mu\nu} = 0$. This leads to
\begin{eqnarray} \label{Tmatch}
 \hat T^{\mu\nu} &\to&  \frac{g_\perp^{\mu\nu}}{Q} \Big( \sum_i O_{1}^{(i)} 
   + \frac{O_{1}^g}{Q} \Big) + \frac{(n^\mu\!+\!\bn^\mu)(n^\nu\!+\!\bn^\nu)}{Q}
   \Big( \sum_i O_{2}^{(i)} + \frac{O_{2}^g}{Q} \Big) \,,
\end{eqnarray}
where
\begin{eqnarray} \label{disop}
  O_j^{(i)} &=& \bar{\xi}^{(i)}_{n,p'} W\, \frac{\bnslash}{2}\, 
      C^{(i)}_j(\bnP_+,\bnP_-)\, W^\dagger \xi^{(i)}_{n,p} \,, \nn\\
  O_j^g &=& \bn_\mu \bn_\nu {\rm tr} \big[ W^\dagger (G_n)^{\mu\lambda} W 
      \: C^g_j(\bnP_+,\bnP_-)\: W^\dagger (G_n)_{\lambda}^{\:\nu}\, W \big],
\end{eqnarray}
where $i$ labels the flavor of the fermions and $ig\, G_n^{\mu\lambda}= [i{\cal
D}_n^\mu + gA_{n,q}^\mu, i{\cal D}_n^\lambda+gA_{n,q'}^\lambda ]$. The Wilson
coefficients are dimensionless functions of $\bnP_+$, $\bnP_-$, $Q$, and $\mu$.
As in previous sections we can separate the hard coefficients from the long
distance operators by introducing trivial convolutions. This gives
\begin{eqnarray} \label{disop1}
  O_j^{(i)} &=& \int\!\!{d\omega_1\,d\omega_2} \,
   C^{(i)}_j(\omega_+,\omega_-)\: \big[ \bCH{n}{\omega_1}^{(i)} 
   \frac{\bnslash}{2} \CH{n}{\omega_2}^{(i)} \big] \,,\nn \\
  O_j^g &=& -\int\!\!{d\omega_1\,d\omega_2}\, C^{g}_j(\omega_+,\omega_-)\:  
   {\rm tr}\big[ B_{n,\omega_1}^{\mu} B^{n,\omega_2}_\mu 
  \big] \,,
\end{eqnarray}
where $\omega_\pm = \omega_1 \pm \omega_2$, and $B^\mu_{n,\omega}\equiv\bn_\nu
({\cal G}_{n,\omega})^{\nu\mu}$ with $({\cal G}_{n,\omega} )^{\mu\lambda}$
defined in Eq.~(\ref{covG}).  Next we factor the coupling of usoft gluons
from the collinear fields using the field redefinitions in Eq.~(\ref{def0}). The
operator $O_j^{(i)}$ has the structure in Eq.~(\ref{decoup}) so the $Y$'s cancel
trivially, while for $O_j^g$ we find
\begin{eqnarray}
  B_n^\mu = Y_n B_n^{\mu(0)} Y_n^\dagger\,,
\end{eqnarray} 
and the factors of $Y$ cancel in the trace. It is easy to see that soft gluons
also decouple using Eq.~(\ref{Sadd}) or by noting that there is no non-trivial
soft gauge invariant way of adding soft Wilson lines $S_n$ to $O_j^{(i)}$ or
$O_j^g$.  Under charge conjugation the full theory electromagnetic current
$J_\mu\to -J_\mu$ and therefore the operator $\hat T_{\mu\nu}\to \hat
T_{\mu\nu}$. This implies relations for the effective theory Wilson coefficients
since the operators $O_j^{(i)}$ must also respect this symmetry. Thus charge
conjugation gives
\begin{eqnarray} \label{DISchg}
  \int\!\!{d\omega_1\, d\omega_2}\, C_j^{(i)}(\omega_+,\omega_-)\ 
  \bCH{n}{\omega_1}^{(i)} {\bnslash}  \CH{n}{\omega_2}^{(i)} 
 &\stackrel{\mbox{\tiny C}}{=}& 
  -\int\!\!{d\omega_1\, d\omega_2}\, C_j^{(i)}(\omega_+,\omega_-)\ 
  \bCH{n}{-\omega_2}^{(i)} {\bnslash}  \CH{n}{-\omega_1}^{(i)}\nn\\
 &=& \int\!\!{d\omega_1\,d\omega_2} \big[ - C_j^{(i)}(-\omega_+,\omega_-)\big] \ 
  \bCH{n}{\omega_1}^{(i)} {\bnslash}  \CH{n}{\omega_2}^{(i)} \,.
\end{eqnarray}
In the second line we changed variable $\omega_1\to -\omega_2$ and $\omega_2\to
-\omega_1$ which takes $\omega_+\to -\omega_+$ and $\omega_-\to \omega_-$. Thus,
to all orders in perturbation theory $C_j^{(i)}(-\omega_+,\omega_-) =
-C_j^{(i)}(\omega_+,\omega_-)$.  This relates the Wilson coefficients for quarks
and anti-quarks.  Note that the above results are all independent of the
collinear hadron on which DIS is performed.

Next we take the matrix element between proton states.  Using the definitions of
the nonperturbative matrix elements given in Section~\ref{section_melt}, and
picking out the coefficients of the tensor structures we find that the delta
functions in Eq.~(\ref{pdfqg}) set $\omega_+=\pm 2Q\xi/x$ and $\omega_-=0$.
Since charge conjugation relates negative and positive values of $\omega_+$ only
coefficients, $C_j(\omega_+,0)$, with positive $\omega_+$ are needed in the
formulae for DIS. Therefore we define
\begin{eqnarray}
  H_j(z) \equiv 
  C_j(2Qz,0,Q,\mu) \,,
\end{eqnarray}
where here we have made the dependence on $Q$ and $\mu$ explicit. Combining this
with Eqs.~(\ref{tprod}), (\ref{Tmatch}), (\ref{disop1}), and (\ref{pdfqg}),
gives the final result
\begin{eqnarray}
  T_1(x,Q^2) &=& - \frac{1}{x} \int_0^1 \!\!d\xi \,
  \bigg\{ H^{(i)}_1\Big(\frac{\xi}{x}\Big)\, 
   \big[ f_{i/p}(\xi) + \bar f_{{i}/p}(\xi) \big]
   +\frac{\xi}{2x} H^{g}_1\Big(\frac{\xi}{x}\Big)\, 
   f_{g/p}(\xi) \bigg\}\,, \nn \\
  T_2(x,Q^2) &=& \frac{4x}{Q^2} \int_0^1 \!\!d\xi 
   \bigg\{ \Big[4\,H^{(i)}_2\Big(\frac{\xi }{x}\Big)
   \!-\! H^{(i)}_1\Big(\frac{\xi}{x}\Big) \Big] 
   \big[ f_{i/p}(\xi)+ \bar f_{{i}/p}(\xi) \big]  \nn\\
 && \qquad\qquad\ \ +\frac{\xi}{2x} \Big[4\,H^{g}_2\Big(\frac{\xi }{x}\Big)
   \!-\! H^{g}_1\Big(\frac{\xi}{x}\Big) \Big] f_{g/p}(\xi) \bigg\} \,,
\end{eqnarray}
where a sum over $i$ is implicit. The hadronic tensor components
$W_{1,2}(x,Q^2)= {\rm Im}\: T_{1,2}(x,Q^2)/\pi$ and therefore are determined by
the imaginary part of the Wilson coefficients. The Wilson coefficients are
dimensionless and therefore can only have $\alpha_s(Q)\ln(\mu/Q)$ dependence on
$Q$. This reproduces the Bjorken scaling of the structure functions.

Finally, consider the tree level matching onto the Wilson coefficients shown in
Fig.~\ref{disopefig}.  From these graphs only the quark coefficient functions
$C^{(i)}_{j}$ can be non-zero and we find
\begin{eqnarray}
 && {\rm Im}\, H^{(i)}_1(z) = {-\,Q_i^2}\,\pi\: \delta(z-1)
   \,,  \qquad\quad {\rm Im}\, H^{(i)}_2 = 0 \,,
\end{eqnarray}
where $Q_i$ is the charge of parton $i$. The vanishing of ${\rm Im}\,H^{(i)}_2$
at tree level reproduces the Callan-Gross relation $W_1/W_2= Q^2/(4x^2)$.


\subsection{Drell-Yan,\ $p \bar p \to \ell^+\ell^- X $}

Next we will extend the DIS analysis to the Drell-Yan (DY) process: $p \bar p
\to \ell^+\ell^- X $. Specifically we consider the $Q^2$ distribution, where
$Q^2$ is the invariant mass of the lepton pair. Drell-Yan is more complicated
than DIS because one has two hadrons in the initial state. In the center-of-mass
frame the incoming proton and anti-proton move in opposite lightlike directions,
and to prove factorization we use the fact that collinear modes in different
lightlike directions can only couple to each other in external operators in
SCET. We take the incoming proton to move in the $n^\mu$ direction and the
incoming antiproton to move in the $\bar{n}^\mu$ direction. The hard scales in
DY are $Q^2$ and the invariant mass of the colliding $p \bar p $ pair
$s=(p+\bar{p})^2$.  The lepton pair has an invariant mass $Q^2$, and the
invariant mass of the final hadronic state is 
\begin{eqnarray}
  p_X^2 = Q^2\left(1+\frac{1}{\tau} - \frac{1}{x_1} - \frac{1}{x_2}  \right)\,,
\end{eqnarray}
where
\begin{eqnarray}
  \tau = \frac{Q^2}{s}\,, \qquad x_1 = \frac{Q^2}{2p \cdot q}\,, \qquad 
   x_2 = \frac{Q^2}{2\bar p \cdot q}\,.
\end{eqnarray}
We are interested in the kinematic region where $p_X^2 \sim Q^2$, which implies
that both $x_1$ and $x_2$ are far away from one. As $\tau$ approaches one the
invariant mass becomes too small for the treatment given here to apply. However,
the effective theory can be used to deal with this region as well. It is also
possible to study the $q_\perp$ distribution, but this again requires a
generalization of the discussion given below.

The spin averaged cross section for Drell-Yan is
\begin{eqnarray} \label{sDY}
 d\sigma  = \frac{32\pi^2 \alpha^2}{Q^4\,s}\:
  L_{\mu \nu}\,W^{\mu \nu}\frac{d^3k_1}{(2\pi)^3(2k_1^0)}
  \frac{d^3k_2}{(2\pi)^3 (2k_2^0)} \label{DY} \,,
\end{eqnarray}
where 
\begin{eqnarray}
 W^{\mu\nu} &=& \frac{1}{4}
 \sum_{{\rm spins}}\sum_X (2\pi)^4 \delta^{(4)}(p+\bar{p}-q-p_X)
 \langle p \bar{p} | J^\mu(0) | X \rangle \langle X| J^\nu(0)|p \bar{p} \rangle
 \nonumber\\
&=& \frac{1}{4} \sum_{{\rm spins}} \int d^4 x\,  e^{-iq\cdot x} 
 \langle p \bar{p} | J^\mu(x) \, J^\nu(0)|p \bar{p} \rangle \,.
\end{eqnarray}
The sum over spins refers to the initial hadron spins (the sum over final hadron
spins is included in the sum over $X$). Integrating Eq.~(\ref{sDY}) over the
emission angles of the final leptons one obtains
\begin{eqnarray}\label{dyqsqrdist}
 \frac{d \sigma}{d Q^2} = \frac{2\alpha^2}{3\, Q^2\,s}\ \frac{1}{4} 
 \sum_{\rm spins}\: \langle p\bar p |\, \hat W\, | p\bar p \rangle  \,, 
\end{eqnarray} 
where we have neglected the lepton masses and defined the operator
\begin{eqnarray} \label{What}
 \hat W(\tau,Q^2) = - \int\!\! \frac{d^4q}{(2\pi)^3}\: \theta(q^0)
   \delta(q^2\!-\!Q^2) \int\!\! d^4x\: e^{-iq\cdot x}\, J^\mu(x)\, J_\mu(0) \,.
\end{eqnarray}
As we discussed above in the region of phase space under consideration
$p_X^2\sim Q^2$, so these hard fluctuations can be integrated out. Operationally
this means we match $\hat W$ onto local operators in the effective theory. We
would like to show that the minimal set of order $\lambda^4$ operators that
contribute to Drell-Yan are
\begin{eqnarray}\label{dymatching}
 \hat W \to &  &  
 \frac{1}{Q^2} \int\!\!{d\omega_i}\: C_{qq}(\omega_i,Q) 
 \big[\bCH{n}{\omega_1}^{\,(i)} \:{\bnslash}\: \CH{n}{\omega_2}^{\,(i)}\big]
 \big[\bCH{\bn}{\omega_3}^{\,(i)} \:{\nslash}\: \CH{\bn}{\omega_4}^{\,(i)}\big]
 \, \nn \\[3pt]
&-& \frac{1}{Q^3} \int\!\!{d\omega_i}\:  C_{qg}(\omega_i,Q)\,  
 \big[\bCH{n}{\omega_1}^{\,(i)} \:{\bnslash}\: \CH{n}{\omega_2}^{\,(i)}\big]
 {\rm Tr}\big[ {\cal B}_{\bn,\omega_3}^{\beta}  
 {\cal B}^{\bn,\omega_4}_{\beta} \big] \, \nn \\[3pt]
&-& \frac{1}{Q^3} \int\!\!{d\omega_i}\:  C_{gq}(\omega_i,Q)
 {\rm Tr}\big[ {\cal B}_{n,\omega_1}^\nu {\cal B}^{n,\omega_2}_{\nu} \big]
 \big[ \bCH{\bn}{\omega_3}^{\,(i)} \:{\nslash}\: \CH{\bn}{\omega_4}^{\,(i)}\big] 
 \,\nn \\[3pt]
&+& \frac{1}{Q^4} \int\!\!{d\omega_i}\: C_{gg}(\omega_i,Q)\,
 {\rm Tr}\big[ {B}_{n,\omega_1}^\nu {B}^{n,\omega_2}_{\nu} \big]
  {\rm Tr}\big[ {\cal B}_{\bn,\omega_3}^{\beta}  
 {\cal B}^{\bn,\omega_4}_{\beta} \big] \,,
\end{eqnarray}
where the powers of $Q$ are included to make the coefficients dimensionless.
The operators displayed in Eq.~(\ref{dymatching}) are just products of the
operators that occured in DIS, so for these terms the decoupling of soft and
usoft gluons occurs in a straightforward manner.  To show that the operators in
Eq.~(\ref{dymatching}) are the most general set needed we must show that all
other operators that are order $\lambda^4$ either reduce to these or vanish
between the matrix elements in Eq.~(\ref{dyqsqrdist}). For instance, $\lambda^4$
operators also exist where a $B_{n,\omega}^\nu$ field is contracted with a
$B_{\bn,\omega'}^\mu$ field, or the color structures of the operators in
Eq.~(\ref{dymatching}) could be arranged in a different way.

We now give a general argument for why we can always rewrite an arbitrary
operator in the form of Eq.~(\ref{dymatching}) or show that it does not
contribute to DY. All operators relevant for DY contain four order $\lambda$
collinear fields chosen from $\xi_{n,p}$, $\xi_{\bn,p}$, $B_{n,p}^\mu$, or
$B_{\bn,p}^\mu$. Furthemore, two must move in direction $n$ and two in the
direction $\bn$ (other possibilities end up vanishing by baryon number
conservation or because they involve a set of fields between physical states
that can not possibly form a color singlet operator).  For operators with 4
quark fields, Fierz transformations can always be made to arrange the fields
such that those in the same direction sit in the same bilinear. Using as an
example the operator with two collinear quarks in the $n$ directions and two
gluons in the $\bn$ direction and leaving out the soft Wilson lines for the
moment, the most general matrix element is
\begin{eqnarray}\label{generalDY}
 \langle p_n \bar{p}_\bn |\, \bCH{n}{\omega_1}^{a,\alpha} 
  \CH{n}{\omega_2}^{b,\beta} \, {\cal B}_{\bn,\omega_3}^{A,\mu} 
  {\cal B}_{\bn,\omega_4}^{B,\nu} \, |p_n \bar{p}_\bn \rangle \; 
  \Delta_{\mu\nu;\alpha\beta}^{ab;AB} \,,
\end{eqnarray}
where $a,b$ are quark colors, $A,B$ are gluon colors, and $\alpha,\beta$ are
spinor indices for the quarks. $\Delta_{\mu\nu;\alpha\beta}^{ab;AB}$ is some
tensor that connects the indices in an arbitrary way. In the contraction of
$a,b$ and $A,B$ there are two possible ways to make an overall color singlet,
one where both the quarks and gluons are in a color singlet, and another where
both the quarks and gluons are in a color octet. We will discuss both of these
possibilities in turn.

\begin{figure}[t!]
  \includegraphics[width=3.5in]{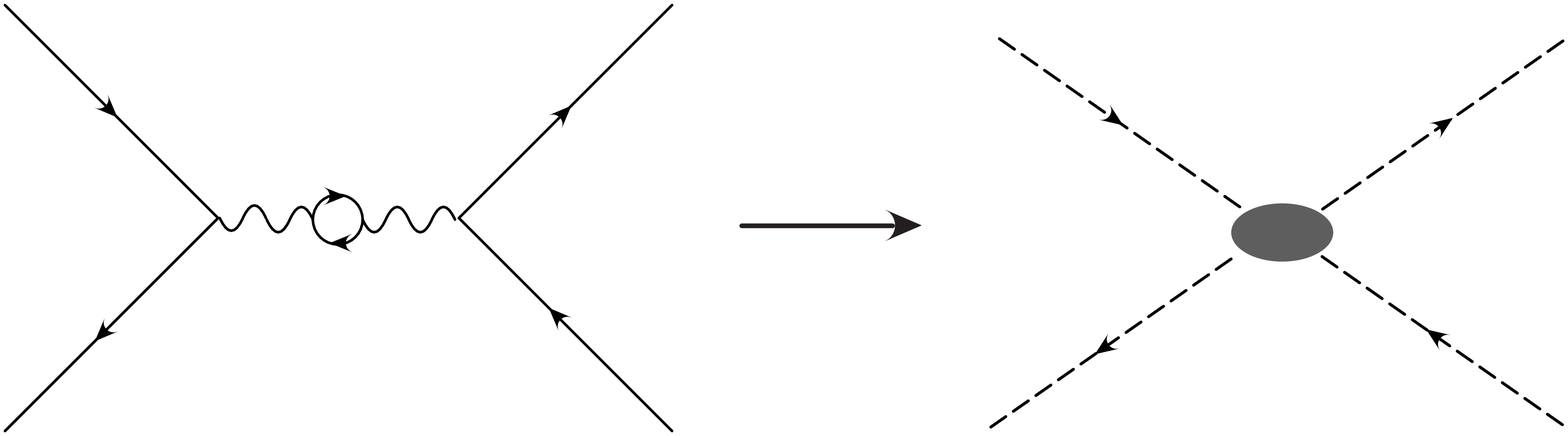}
{\caption{Tree level matching onto the operators in Drell-Yan.
\label{dyfig}}}
\end{figure}
In the color singlet case, including the soft and ultrasoft Wilson lines is
trivial, since using Eqs.~(\ref{Sadd}), (\ref{Yfd2}), and (\ref{eom2}) we see
that they cancel due to unitarity/orthogonality of the various Wilson lines in
the fundamental/adjoint representations. Thus, there are no soft, usoft, or
collinear interactions that connect the $n$ and the $\bn$ fields. As in previous
sections this leads to a factorization of the matrix element in
Eq.~(\ref{generalDY}), namely
\begin{eqnarray}
\langle p_n  |\, \bCH{n}{\omega_1}^{(0)a,\alpha} \CH{n}{\omega_2}^{(0)b,\beta} 
 \, |p_n \rangle \, \langle \bar{p}_\bn | {\cal B}_{\bn,\omega_3}^{(0)A,\mu}
{\cal B}_{\bn,\omega_4}^{(0)B,\nu} \, | \bar{p}_\bn \rangle
\; \Delta_{\mu\nu;\alpha\beta}^{ab;AB} \,.
\end{eqnarray}
Since the proton spins are summed over, we can write (with the help of
Eq.~(\ref{tracefomulae}))
\begin{eqnarray}\label{quarkmatrix}
 \langle p_n  |\, \bCH{n}{\omega_1}^{(0)a,\alpha} \CH{n}{\omega_2}^{(0)b,\beta} 
 \, |p_n \rangle \propto \; \delta^{ab} (\nslash)_{\alpha\beta}\:
\langle {p}_n  |\, \bCH{n}{\omega_1}^{(0)c,\gamma}\bnslash
\CH{n}{\omega_2}^{(0)c,\gamma} \, | {p}_n \rangle \,,
\end{eqnarray}
so that spin and color are summed over in the matrix element. Similarly the
antiproton matrix element can be simplified to
\begin{eqnarray}\label{gluonmatrix}
\langle \bar{p}_\bn | \, {\cal B}_{\bn,\omega_3}^{(0)A,\mu}
{\cal B}_{\bn,\omega_4}^{(0)B,\nu} \, | \bar{p}_\bn\rangle\propto \; \delta^{AB}
  g^{\mu\nu}_\perp
\langle \bar{p}_\bn | \, {\rm Tr}\:\big[ {\cal B}_{\bn,\omega_3}^{(0)\alpha}
         {\cal B}^{(0)\bn,\omega_4}_{\alpha} \big] \, | \bar{p}_\bn \rangle\,.
\end{eqnarray}
Here we used the fact that the matrix element is symmetric in $\mu$ and $\nu$,
and that only the perpendicular index $\mu$ of the field $B^\mu$ is order
$\lambda$.  Using Eqs.~(\ref{quarkmatrix}) and (\ref{gluonmatrix}) the original
matrix element in Eq.~(\ref{generalDY}) can be written as
\begin{eqnarray} \label{endDYeg}
  && \langle p_n \bar{p}_\bn |\, \bCH{n}{\omega_1}^{a,\alpha}
  \CH{n}{\omega_2}^{b,\beta}
  \, {\cal B}_{\bn,\omega_3}^{A,\mu} {\cal B}_{\bn,\omega_4}^{B,\nu}
  \, |p_n \bar{p}_\bn \rangle \; \Delta_{\mu\nu;\alpha\beta}^{ab;AB}
  \\[4pt]
&& \qquad\qquad
  \propto \; {\rm Tr}[\Delta^{\:\mu}_\mu\, \nslash]\,
\langle \bar{p}_\bn | \, {\rm Tr}\:\big[ {\cal B}_{\bn,\omega_3}^{(0)\nu}
         {\cal B}^{(0)\bn,\omega_4}_{\nu} \big] \, | \bar{p}_\bn \rangle\,
\langle p_n  |\, \bCH{n}{\omega_1}^{(0)} \bnslash
\CH{n}{\omega_2}^{(0)} \, |p_n \rangle \,,\nn
\end{eqnarray}
where the trace of $\Delta$ is over spin and color, and just gives an overall
constant. The final result in Eq.~(\ref{endDYeg}) is identical to the matrix
element of the second operator in Eq.~(\ref{dymatching}).  

If each of the $n$ and $\bn$ field bilinears involve color octet structures,
then the soft and usoft Wilson lines don't cancel, since they don't commute with
the SU(3) generators. However, one can use the color identity
\begin{eqnarray} \label{cident}
 Y_n^\dagger S_n^\dagger {\cal T}^x S_\bn Y_\bn \otimes Y_\bn^\dagger 
 S_\bn^\dagger {\cal T}^x S_n Y_n = {\cal T}^x \otimes S_n Y_n Y_\bn^\dagger 
 S_\bn^\dagger {\cal T}^x S_n Y_n Y_\bn^\dagger S_\bn^\dagger
\end{eqnarray}
where each ${\cal T}^x$, $S$, and $Y$ factor is in the appropriate
representation of the color group.  Eq.~(\ref{cident}) moves all the soft and
ultrasoft interactions between either the $n$ or the $\bn$ collinear
fields. Thus, again the fields in one bilinear can not be contracted with fields
in the other bilinear and the matrix element factors. However this time the
factored matrix element vanishes. For the example discussed above,
\begin{eqnarray}\label{quarkmatrixoctet}
  \langle p_n  |\, \bCH{n}{\omega}^{(0)\alpha}\: 
  T^C\CH{n}{\omega'}^{(0)\beta} \, |p_n \rangle= 0\,,
\end{eqnarray}
since an color octet operator vanishes between color singlet states. The same
holds true for the matrix element of an octet gluon operator. 

An identical proof of decoupling goes through for the case of 4 quarks, where we
again either have two color singlet or two color octet $n$ and $\bn$
bilinears. With 4 gluon fields we can either have the two $n$ and two $\bn$
fields coupled as singlets, or coupled in the same higher representation (an
{\bf 8}, \{{\bf 10}, $\overline{\mathbf{ 10}}$\}, or {\bf 27}). In the latter
case the matrix element between color singlet states still vanishes so the proof
for the 4 gluon operators also goes through in an identical way.

Thus we have shown that the matrix element of an operator with an arbitrary
contraction of indices either vanishes or can be written in terms of a product
of a matrix element which is related to a proton pdf and a matrix element which
is related to an antiproton pdf as in the example in Eq.~(\ref{endDYeg}). This
is the result we want. To see how the final formulae are derived note that we can
write the matrix element of Eq.~(\ref{dymatching}) in the form of a convolution
\begin{eqnarray}  \label{dyconv}
 \frac{1}{4}\sum_{{\rm spins}} \langle p_n \bar{p}_\bn |\, \hat W \, 
 |p_n \bar{p}_\bn \rangle
 &=&  \sum_{a,b} \int\!\!{d\omega_i}\:C_{a,b}(\omega_i)
 \langle p_n |O^a_n(\omega_+, \omega_-)| p_n \rangle
 \langle \bar{p}_\bn |O^b_\bn(\omega'_+, \omega'_-)| \bar{p}_\bn \rangle \,,\nn\\
\end{eqnarray}
where $\omega_\pm = \omega_1\pm \omega_2$ and $\omega_\pm'=\omega_3\pm\omega_4$.
The operators here are the same as in DIS, with $a=(i)$ for the quark operator,
and $a=g$ for the gluon operator
\begin{eqnarray} \label{dyop1}
  O^{(i)}_n(\omega_+, \omega_-) &=& \frac{1}{Q}
\big[ \bCH{n}{\omega_1}^{(i)} \frac{\bnslash}{2} 
   \CH{n}{\omega_2}^{(i)} \big] \,,\nn \\
  O^g_n(\omega_+, \omega_-) &=& -\frac{1}{Q^2}
   {\rm tr}\big[ B_{n,\omega_1}^\mu B^{n,\omega_2}_\mu  \big] \,.
\end{eqnarray} 
Apart from the dependence on the labels, the Wilson coefficients in
Eq.~(\ref{dyconv}) can also depend on the renormalization point $\mu$ and the
kinematic variable $Q$. Using Eq.~(\ref{pdfqg}) we see that the matrix elements
in Eq.~(\ref{dyop1}) set $\omega_-=\omega_-'=0$ and $\omega_+=2\sqrt{s} z_1$,
$\omega_+'=2\sqrt{s} z_2$ where $z_1$ and $z_2$ are the convolution variables.
Since all kinematic variables aside from $Q^2$ are integrated over in
Eq.~(\ref{What}) the only other variable that can appear in the Wilson
coefficient is the center of mass energy which produces the $\ell^+\ell^-$ pair,
namely $4\hat{s} = \omega_+ \omega'_+$. Thus, the Wilson coefficients only
depend on $\omega_+\omega_+'=4s z_1 z_2$.  Defining new coefficients
\begin{eqnarray}
  H^{ab}(z_1 z_2) &=&   C^{ab}(\omega_+\omega_+'=4s z_1 z_2,Q,\mu)
\end{eqnarray}
we can replace the matrix elements in Eq.~(\ref{dyop1}) with parton distribution
functions using Eq.~(\ref{pdfqg}) to obtain:
\begin{eqnarray}
  \frac{1}{4}\sum_{{\rm spins}} \langle p_n\bar p_\bn |\, 
   \hat W\, | p_n\bar p_\bn \rangle 
 &=&  \frac{1}{\tau} \int_0^1 \!dz_1\: dz_2 \, 
  \bigg\{ \!-\!H^{(i)(j)}(-\!z_1 z_2) \Big[ f_{i/p}(z_1)\bar{f}_{j/\bar{p}}(z_2)+
    \bar{f}_{i/p}(z_1) f_{j/\bar{p}}(z_2) \Big]\nn\\[3pt]
 &&\hspace{-1cm}+  H^{(i)(j)}(z_1 z_2) \Big[ f_{i/p}(z_1) f_{j/\bar{p}}(z_2)+
    \bar{f}_{i/p}(z_1) \bar{f}_{j/\bar{p}}(z_2) \Big]\nonumber\\[3pt]
 &&\hspace{-1cm} + \frac{z_2}{2\sqrt{\tau}}\: H^{(i)g}(z_1 z_2) f_{i/p}(z_1)  
     f_{g/\bar{p}}(z_2)- \frac{z_2}{2\sqrt{\tau}}\: H^{(i)g}(-z_1 z_2)
     \bar{f}_{i/p}(z_1) f_{g/\bar{p}}(z_2) \nn\\[3pt]
 &&\hspace{-1cm} + \frac{z_1}{2\sqrt{\tau}}\: H^{g(j)}(z_1 z_2) f_{g/p}(z_1) 
    f_{j/\bar{p}}(z_2) -\frac{z_1}{2\sqrt{\tau}}\: H^{g(j)}(-z_1 z_2) 
    f_{g/p}(z_1)\bar{f}_{j/\bar{p}}(z_2) \nn\\[3pt]
 &&\hspace{-1cm} + \frac{z_1 z_2}{4\tau}\: H^{gg}(z_1 z_2) f_{g/p}(z_1) 
  f_{g/\bar{p}}(z_2) \bigg\} \,.
\end{eqnarray}
This is the final convolution formulae for Drell-Yan and is valid to all orders
in $\alpha_s$ and leading order in the power expansion. 
At tree level the matching calculation shown in Fig.~\ref{dyfig} yields zero for
all the Wilson coefficients except
\begin{eqnarray}
  H^{(i)(i)}(-z_1 z_2) =-\frac{2\pi\tau}{3}\: {Q_i^2} \,\delta(\tau - z_1 z_2) 
  \,,
\end{eqnarray}
where $Q_i$ is the charge of parton $i$. The coefficients $H^{(i)g}(\pm z_1
z_2)$ and $H^{g(j)}(\pm z_1 z_2)$ start at order $\alpha_s(Q)$, while 
$H^{(i)(i)}(z_1 z_2)$ and $H^{gg}(z_1 z_2)$ start at order $\alpha_s^2(Q)$.

\subsection{Deeply Virtual Compton Scattering, $\gamma^* p\to \gamma^{(*)} p'$}

Next we examine deeply virtual Compton scattering (DVCS). To be more precise we
examine the exclusive reaction $\gamma^* p \to \gamma^{(*)} p$, where the
incoming photon is highly virtual, the final photon is either off-shell or real,
and the incoming and outgoing protons have different momenta. The reason we have
included this process in the inclusive section is that DVCS has the remarkable
property that the nonperturbative physics is described by a so called
non-forward parton distribution function (NFPDF). The NFPDF is a more general
distribution function that reduces to the standard parton distribution functions
(familiar from DIS) for some values of the momentum fraction, and behaves like a
lightcone wavefunction (familiar from the pion examples) for other
values. Deeply virtual Compton scattering was first studied in perturbative QCD
in Refs~\cite{Ji:1996ek,Ji:1996nm,Radyushkin:1996nd}, and proofs of
factorization to all orders in perturbation theory were later presented in
Refs~\cite{Collins:1998be,Ji:1998xh}. In addition properties of NFPDFs were
studied in Ref.~\cite{Radyushkin:1997ki}.  Here we present a proof of
factorization for DVCS based on SCET.

As with the previous proofs it is important to understand the kinematics of the
process. We take the incoming proton and photon momenta to be $p$ and $q$
respectively, with $x = Q^2/(2 p\cdot q)$ and $q^2 = - Q^2 \gg \Lambda_{{\rm
QCD}}^2$. The outgoing proton and photon momentum are $p'$ and $q'$
respectively, with $0 \geq q^{\prime 2} \geq -Q^2$. It is convenient to define a
parameter $\zeta\equiv 1-\bn\mcdot p^\prime/\bn\mcdot p$, which measures the
change to the proton's large momentum.  Working in the Breit frame and
neglecting contributions that are $\ll \Lambda_{{\rm QCD}}^2/Q$ we have:
\vspace{0.1cm}
\begin{eqnarray} \label{DVCSkin}
\begin{tabular}{llll}
 &  &  \mbox{\normalsize Label Momenta}
  & \mbox{\normalsize Residual Momenta}\\ \hline \\[-12pt]
 & $q^\mu=$\hspace{0.8cm} & $\mfrac{Q}{2} (\bn^\mu\!-\! n^\mu)$ 
  & $+0$ \\[5pt]
 & $p^\mu=$ & $\mfrac{Q}{2 x} \, n^\mu$ 
  & $+\mfrac{x}{2Q}\, m_p^2(n^\mu\!+\!\bn^\mu)$ \\[5pt]
 & $p^{\prime\mu}=$  & $\mfrac{Q}{2 x}  (1\!-\!\zeta) \, n^\mu 
   \!+\! p^{\prime\mu}_\perp$ & $+\mfrac{x}{2Q} \big[ m_p^2(1\!-\! \zeta)n^\mu 
  \!+\! \{m_p^2(1\!+\!\zeta)\!-\! t\}\bn^\mu\big]$\hspace{0.3cm}  \\[5pt]
 & $q^{\prime\mu}=$ & $\mfrac{Q}{2} \big(\mfrac{\zeta}{x}-\! 1 \big)
  \, n^\mu + \mfrac{Q}{2}\, \bn^\mu  \!-\! p_\perp^{\prime\mu}$\hspace{0.7cm} 
  & $+\mfrac{x}{2Q} \big[ \zeta m_p^2 n^\mu 
  + (t\!-\!\zeta m_p^2)\bn^\mu\big]$ \\[5pt] \hline
\end{tabular} \hspace{0.5cm} \vspace{0.2cm}
\end{eqnarray}
Here the label momenta are order $Q$ or $Q\lambda$, while the residual momenta
are order $Q\lambda^2$ and depend on $p^2=m_p^2$ and
$t=(p^\prime-p)^2=(p_\perp^{\prime 2}\!-\!\zeta^2 m_p^2)/(1\!-\!\zeta)$, which
are both $\sim \Lambda_{\rm QCD}^2$. The invariant mass of the intermediate
hadronic state is $(p+q)^2 \approx Q^2(1-x)/x$ just like DIS, so for $1-x \gg
\Lambda_{{\rm QCD}}/Q$ the intermediate state can be integrated out.

We will proceed in a manner analogous to the analysis for DIS. The amplitude (up
to an overall momentum conserving $\delta$-function) is given by a time ordered
product of currents:
\begin{eqnarray} 
 T_{\mu\nu}(p,q,q') &=& \langle p',\sigma' | \hat{T}_{\mu\nu}(q,q') 
 | p,\sigma \rangle \nonumber \\
 \hat{T}_{\mu\nu}(q,q')  &=& i \int d^4z \, e^{i(q+q')\cdot z/2}\,
 T\big[J_\mu(-z/2) J_\nu(z/2)\big] \,.
\end{eqnarray}
This time ordered product is contracted with a lepton tensor to obtain the
amplitude. Now current conservation requires $q^\mu T_{\mu\nu}= q^{\prime\nu}
T_{\mu\nu}=0$, however the DVCS $T_{\mu\nu}$ is not symmetric under $\mu
\leftrightarrow \nu$. For electromagnetic currents $J_\mu$ we have
\begin{eqnarray} \label{TDVCS}
  T_{\mu\nu} &=& 
  -\Big( g_{\mu\nu} -\frac{q^\prime_\mu q_\nu}{q\mcdot q'} \Big)\: T_1
  + \Big( p_\mu - \frac{q^\prime_\mu\,p\mcdot q}{q\mcdot q'} \Big)
    \Big( p_\nu - \frac{q_\nu\,p\mcdot q'}{q\mcdot q'} \Big)\: T_2 \nn\\
 && + \ell_\mu \ell^{\prime}_\nu\: T_3 
    + \ell_\mu \Big( p_\nu - \frac{q_\nu\,p\mcdot q'}{q\mcdot q'} \Big)\: T_4
    + \Big( p_\mu - \frac{q^\prime_\mu\,p\mcdot q}{q\mcdot q'} \Big)\ell_\nu
      \: T_5
    + \ldots\,,
\end{eqnarray}
where the functions $T_i=T_i(x,\zeta,Q^2,t)$, and the vectors $\ell_\mu\equiv
q^\prime_\mu-q_\mu + p_\mu (q^2\!-\!q^{\prime 2}\!+\!t) /(2p\cdot q)$ and
$\ell_\mu^\prime\equiv q_\mu-q^\prime_\mu + p_\mu (q^{\prime 2}\!-\!q^{2}\!+\!t)
/(2p\cdot q^\prime)$ are defined so that $q\cdot\ell = q^\prime\cdot\ell^\prime
= 0$.  In Eq.~(\ref{TDVCS}) and below the ellipses denote spin dependent
terms. For simplicity we will show how factorization is achieved for the spin
independent contributions shown in Eq.~(\ref{TDVCS}) with the understanding that
it is no more difficult to also include the other terms.

It is convenient to define a parameter $0\leq\alpha\leq 1$, by $q^{\prime
2}\equiv -\alpha Q^2$. The DIS hadronic time-ordered product is obtained in the
limit $p' \to p$, where $\alpha \to 1$ and $\zeta \to 0$. From
Eq.~(\ref{DVCSkin}) we see that
\begin{eqnarray}
 \zeta=x(1-\alpha)+{\cal O}\Big(\frac{t}{Q^2},\frac{m_p^2}{Q^2}\Big)\,,
\end{eqnarray}
so these parameters are not independent.  Since the intermediate hadronic state
has invariant mass ${\cal O}(Q^2)$ we can match $\hat{T}_{\mu\nu}$ onto
operators in SCET. Requiring $q^\mu \hat{T}_{\mu\nu}=0$ and $q^{\prime\nu}
\hat{T}_{\mu\nu}=0$ for the order $Q$ label momenta leads to
\begin{eqnarray}
 \hat{T}_{\mu\nu} \to \frac{g^{\mu\nu}_\perp}{Q} 
 \Big(O^{(i)}_1+\frac{O^{g}_1}{Q} \Big)+
 \frac{1}{Q} \left( n^\mu + \bn^\mu \right)
 \left( \alpha n^\nu + \bn^\nu \right) 
 \Big(O^{(i)}_2+\frac{O^{g}_2}{Q} \Big) +\ldots \,,
\end{eqnarray}
where the ellipses are spin dependent terms and the displayed operators are 
\begin{eqnarray} \label{dvcsop}
  O_j^{(i)} &=& \bar{\xi}^{(i)}_{n,p'} W\, \frac{\bnslash}{2}\, 
      C^{(i)}_j(\bnP_+,\bnP_-)\, W^\dagger \xi^{(i)}_{n,p} \,, \nn\\
  O_j^g &=& \bn_\mu \bn_\nu\: {\rm tr} \big[ W^\dagger (G_n)^{\mu\lambda} W 
      \: C^g_j(\bnP_+,\bnP_-)\: W^\dagger (G_n)_{\lambda}^{\:\nu}\, W \big] \,.
\end{eqnarray}
We have suppressed the dependence of the Wilson coefficients $C(\bnP_+,\bnP_-)$
on $Q$, $\alpha$, and $\mu$. The form of the operators in Eq.~(\ref{dvcsop})
looks the same as the DIS operators given in Eq.~(\ref{disop}), however the
operators here are more general because the Wilson coefficients depend on
$\alpha$. In the limit $\alpha \to 1$ the DVCS operators reduce to the DIS
operators. However, since the field structure of the DVCS operators is identical
to DIS several results follow immediately. For instance, the steps which
factorize soft and usoft gluons and leave fields with superscript $(0)$ are the
same and are not repeated here,
\begin{eqnarray} \label{dvcsop2}
   O_j^{(i)} &=& \bCH{n}{\omega}^{\,(0)(i)} \frac{\bnslash}{2}\, 
      C^{(i)}_j(\bnP_+,\bnP_-)\, \CH{n}{\omega'}^{\,(0)(i)} \,, \nn\\
  O_j^g &=& -{\rm tr} \Big[ B_{n,\omega}^{(0)\mu}
      \: C^g_j(\bnP_+,\bnP_-)\: (B_{n,\omega'}^{(0)})_\mu \Big] \,.
\end{eqnarray}
The restrictions on the DVCS Wilson coefficients from charge conjugation are the
same as in Eq.~(\ref{DISchg}), $C_j(\bnP_+,\bnP_-)=-C_j(-\bnP_+,\bnP_-)$,
however because $p\ne p'$ this is not simply a relation between quark and
anti-quark Wilson coefficients. The way in which DVCS is unique is that the
matrix elements involve nucleon states with different momenta. This is what
leads to results in terms of non-forward parton distribution functions.

The definition of the NFPDFs are given in Eq.~(\ref{nfpdfqg}), and can be used
along with the relations above to obtain expressions for the $T_i$ in terms of
the NFPDFs. Before we give this result we note that the Wilson coefficients
depend on the operators $\bnP_\pm$ which become the variables $\omega_\pm$ after
introducing trivial convolutions and the $\chi_{n,\omega_i}$ fields in
Eq.~(\ref{CH1}). The delta functions in Eq.~(\ref{nfpdfqg}) then set $\omega_-=
-Q \zeta / x$ and $\omega_+= \pm Q (2 \xi - \zeta)/x$, where $\xi$ is the
convolution variable. Note that $\zeta/x = 1-\alpha$, and just like DIS it is
the combination $\xi/x$ which appears. Since charge conjugation relates the 
Wilson coefficients for $\omega_+>0$ and $\omega_+<0$ it is convenient to define
\begin{eqnarray} \label{Hdvcs}
  {\cal H}_j(\xi/x) \equiv 
  C_j\big(Q(2\xi/x\!- \!1\!+\!\alpha),Q(\alpha\!-\!1),Q,\alpha,\mu\big) \,,
\end{eqnarray}
where in the last three arguements we have made the dependence on $Q$, $\alpha$,
and $\mu$ explicit.  Combining Eqs.~(\ref{nfpdfqg}), (\ref{dvcsop2}), and
(\ref{Hdvcs}) then gives
\begin{eqnarray}\label{tmatch}
  T_1 &=&  -\frac{e(\sigma^\prime,\sigma)}{2Q}
  \int_0^1 \!\!d\xi \,
  \bigg\{ {\cal H}^{(i)}_1 \! \Big(\frac{\xi}{x}\Big)\:
   \Big[ {\cal F}^{i}_\zeta(\xi;t) + \overline{{\cal F}}^{i}_\zeta(\xi;t) \Big] 
  + \frac{1}{2x} {\cal H}^{g}_1 \Big(\frac{\xi}{x}\Big)\: 
  {\cal F}^{g}_\zeta(\xi;t) \bigg\}  +\ldots \,, \nn \\[3pt]
  T_2 &=& \frac{x^2(1+\alpha)}{Q^3}\: e(\sigma^\prime,\sigma)
  \int_0^1 \!\!d\xi  \bigg\{ \Big[ 2(1+\alpha) {\cal H}^{(i)}_2 \!
  \Big(\frac{\xi}{x}\Big) \!-\! {\cal H}^{(i)}_1 \! \Big(\frac{\xi}{x}\Big)
  \Big] \Big[{\cal F}^{i}_\zeta(\xi;t)+\overline{{\cal F}}^{i}_\zeta(\xi;t)\Big]
  \nn  \\
 && + \frac{1}{2x} \Big[ 2(1+\alpha)\,
   {\cal H}^{g}_2 \Big(\frac{\xi}{x}\Big)
   \!-\! {\cal H}^{g}_1  \Big(\frac{\xi}{x}\Big) \Big] 
   {\cal F}^{g}_\zeta(\xi;t) \bigg\} +\ldots  \,, \nn\\[3pt]
 T_3 &=& 0\,, \qquad T_4 =0\,, \qquad T_5 =0 \,,
\end{eqnarray}
which are the final convolution results valid to all orders in $\alpha_s$ and
leading order in the power expansion.  The structure functions $T_{3,4,5}$
vanish since the vectors $\ell^\mu=\ell^{\prime\mu}=0$ at leading order in the
power expansion. The terms with ellipses are for the spin flip terms and involve
the NFPDF ${\cal K}$ defined in Eq.~(\ref{nfpdfqg}). The results for these terms
have a similar form to those in Eq.~(\ref{tmatch}).

Finally we match at tree level. The tree level diagram in QCD is the same as in
Fig.~\ref{disopefig} except the outgoing photon and proton have momenta
$q^\prime$ and $p^\prime$ respectively. Only the quark Wilson coefficients are
nonzero at tree level. We find
\begin{eqnarray}
C_1^{(i)}(\omega_+,\omega_-,Q,\alpha) &=& e^2 Q^2_i
   \bigg( \frac{2 Q}{2 Q +\omega_+ - \omega_-} - 
  \frac{2 Q}{2Q-\omega_+ - \omega_-} \bigg) 
\\
C_2^{(i)}(\omega_+,\omega_-,Q,\alpha) &=& 0 \,,
\nn
\end{eqnarray}
which gives
\begin{eqnarray}
{\cal H}_1^{(i)}\left(\frac{\xi}{x}\right) &=& -e^2 Q^2_i
   \bigg( \frac{1}{1-\xi/x} - \frac{1}{1+(\xi-\zeta)/x} 
   \bigg) \\
 {\cal H}_2^{(i)}\left(\frac{\xi}{x}\right) &=& 0 \,.\nn
\end{eqnarray}
Since ${\cal H}_2^{(i)}=0$ at tree level, DVCS also obeys a Callan-Gross
relation.

\section{Conclusion}\label{section_conclusion}

What we hope we have demonstrated here is the power of effective field
techniques in the context of factorization for hard scattering processes. The
explicit separation of modes and the implementation of gauge invariance for
these modes greatly simplifies seemingly complex problems. What is normally
accomplished by diagramatic Ward identities and induction techniques now falls
out as a consequence of the gauge symmetry of operators in a low energy
soft-collinear effective theory.

As we have emphasized the factorization formulae derived in this paper are not
new. The purpose here was simply to extend the formalism introduced in
\cite{bfl,bfps,cbis,bpssoft,bps} to cases with back-to-back collinear particles,
and apply these ideas to more general processes than previously considered.
Furthermore, the factorization proofs presented are perhaps simpler than those
previously given (certainly they are more concise).  We believe that within the
confines of the SCET more difficult, and unresolved problems can be addressed,
such as power corrections in cases without an OPE, and proofs of factorization
for more complex processes.

\begin{acknowledgments}
This work was supported in part by the Department of Energy under the grant
DOE-FG03-97ER40546.
\end{acknowledgments}

\newpage

\appendix

\section{Factorization of soft and collinear $n$ \& $\bn$ modes}
\label{app_fact}

This Appendix discusses the simultaneous factorization of the soft
$(\lambda,\lambda,\lambda)$ modes, $n$-collinear $(\lambda^2,1,\lambda)$ modes,
and $\bn$-collinear $(1,\lambda^2,\lambda)$ modes. These three classes of modes
can not interact with each other in a local manner and therefore do not couple
through the SCET Lagrangian. However, they can couple in a gauge invariant way
through external operators and currents. These interactions in currents are
built up by integrating out offshell fluctuations with $p^2\gg (Q\lambda)^2$.
For the special case of factorization of soft from $n$-collinear modes this was
shown in detail in the Appendix of Ref.~\cite{bpssoft}.  There it was shown that
integrating out certain modes with offshellness $p^2\sim Q^2\lambda$ causes the
Wilson lines $W_n$ and $S_n$ to appear in operators in a gauge invariant
way. Here we will extend this approach to the factorization of modes for cases
involving two classes of collinear particles. For simplicity we restrict
ourselves to the case where the original operators involve only collinear quark
or gluon fields. This type of factorization was used for the pion form factor
example discussed in Sec.~III.B and the Drell-Yan process presented in
Sec.~IV.B.

\begin{table}[t!]
\begin{center}
\begin{tabular}{cl|l|c|ccc}
  \multicolumn{2}{c}{Type} & \hspace{0.2cm} Momenta $(+,-,\perp)$\hspace{0.4cm} 
   & \hspace{0.2cm} Fields\hspace{0.2cm}  & \ Wilson lines & \\ \hline
 onshell\hspace{.3cm} & collinear-$n$ & \hspace{0.3cm} 
   $p_1^\mu\sim (\lambda^2,1,\lambda)$ & $\xi_{n}$, $A_{n}^\mu$ & $W_n$ \\
 & collinear-$\bn$ & \hspace{0.3cm} $p_2^\mu\sim (1,\lambda^2,\lambda)$ 
   & $\xi_{\bn}$, $A_{\bn}^\mu$ & $W_\bn$ \\
  & soft &  \hspace{0.3cm} $q^\mu\sim (\lambda,\lambda,\lambda)$ 
   & $q_{s}$, $A_{s}^\mu$ & $S_n$, $S_\bn$ \\ 
  & usoft & \hspace{0.3cm} $k^\mu\sim (\lambda^2,\lambda^2,\lambda^2)$ 
   & $q_{us}$, $A_{us}^\mu$ & $Y_n$, $Y_\bn$ \\[3pt] \hline
 offshell\hspace{.3cm} & $p=p_1+p_2$ & \hspace{0.3cm} $p^\mu \sim (1,1,\lambda)$
   & $\psi_Q$, $A_Q^\mu$ & $X_n$, $X_\bn$ \\
 & $p=p_1+q$ & \hspace{0.3cm}  $p^\mu\sim (\lambda,1,\lambda)$ 
   & $\psi^L_{n}$, $A^{X\mu}_{n}$ & $W_n^X$, $S_n^X$ \\
 & $p=p_2+q$ & \hspace{0.3cm}  $p^\mu\sim (1,\lambda,\lambda)$ &
   $\psi^L_{\bn}$, $A_{\bn}^{X\mu}$ & $W_\bn^X$, $S_\bn^X$ \\ 
 \hline
\end{tabular}
\end{center}
{\caption{Summary of the onshell modes discussed in section~\ref{section_SCET}, 
and the auxiliary fields introduced to represent the offshell fluctuations that 
are integrated out in this appendix.
\label{table_off} }}
\end{table}

The basic idea is to first match onto a Lagrangian with couplings between
onshell and offshell modes that give all order $\lambda^0$ diagrams. The
offshell modes (with $p^2\gg (Q\lambda)^2$) can then be integrated out, so
that all operators are expressed entirely in terms of the onshell degrees of
freedom. In table~\ref{table_off} a summary is given of the three types of
offshell momenta that are induced by adding soft, $n$-collinear, and
$\bn$-collinear momenta. For each type auxiliary quark and gluon fields are
defined, and for convenience momentum labels are suppressed in this section. For
example, the $\psi_Q$ quarks are created by the interaction of a $n$-collinear
quark with an $\bn$-collinear gluon, whereas the $\psi_n^L$ quarks are created
when a collinear quark $\xi_n$ is knocked offshell by a soft gluon. For the
field $\psi_Q$ we write $\psi_Q = \psi^Q_n + \psi_\bn^Q$, where $\psi^Q_n =
\frac{1}{4}{\nslash\bnslash}\, \psi_Q$ and $\psi^Q_\bn =
\frac{1}{4}\bnslash\nslash\, \psi_Q$. Then we have $\nslash \psi_n^Q = \bnslash
\psi_\bn^Q =0$ and $\nslash \psi_n^L =\bnslash \psi_\bn^L =0$.  Various Wilson
lines are also required and are listed in the table. It is convenient to define
a generic Wilson line $L[a,A]$ along direction $a$ with field $A$ by the
solution of
\begin{eqnarray}
  \big( a\mcdot\cP + g\, a\mcdot A\big) L[ a, A] = 0 \,.
\end{eqnarray}
With this notation the on-shell Wilson lines are $W_n=L[\bn,A_n]$,
$W_\bn=L[n,A_\bn]$, $S_n=L[n,A_s]$, and $S_\bn=L[\bn,A_s]$. (Recall that the
subscripts on $W$ and $S$ mean different things.) The Wilson lines involving
offshell fields that we will require are 
\begin{eqnarray}
  X_n &=& L[\bn,A_Q\!+\!A_n^X\!+\!A_n] \,, \qquad 
  X_\bn=L[n,A_Q\!+\!A_\bn^X\!+\!A_\bn] \,,\\
  W_n^X &=& L[\bn,A_n^X\!+\!A_n] \,, \qquad\qquad\!
  W_\bn^X=L[n,A_\bn^X\!+\!A_\bn] \,,\nn \\
  S_n^X &=& L[n,A_n^X\!+\!A_s]\,, \qquad\qquad\  
  S_\bn^X = L[\bn,A_\bn^X\!+\!A_s]  \,.\nn
\end{eqnarray}
Below we discuss the results which allow us to integrate out offshell
fluctuations. The structure of the auxiliary Lagrangians and construction of
their solutions are very similar to the case presented in Ref.~\cite{bpssoft},
to which we refer for a more detailed presentation.

From table~\ref{table_off} we see that adding $n$ and $\bn$-collinear momenta
gives $p^2\sim Q^2$, whereas adding soft and collinear momenta gives $p^2\sim
Q^2\lambda$. Loops that are dominated by offshell momenta only modify Wilson
coefficients and not the infrared physics. Therefore, to determine the structure
of SCET fields in an operator it sufficient to integrate out the offshell fields
at tree level.  For convenience we can integrate out the fluctuations starting
with those with the largest offshellness. Recall that we only wish to consider
offshell propagators connected to external operators. A subtlety for quarks is
that distinct auxiliary fields are needed for the incoming and outgoing offshell
propagators. However, the solution for the outgoing field turns out to be the
conjugate of the incoming field, so to avoid a proliferation of notation we
simply denote the outgoing terms in the Lagrangian by $+\mbox{h.c.}$, and
present a solution for the incoming fields. Finally, note that for the gluon
field $A_Q$ the fields $A_n$, $A_\bn$, $A^X_n$, $A_\bn^X$, and $A_s$ appear as
background fields while for the fields $A_n^X$ and $A_\bn^X$ it is $A_n$,
$A_\bn$, and $A_s$ that appear as background fields.

The terms in the auxiliary Lagrangian involving the $p^2\sim Q^2$ fields are
\begin{eqnarray}\label{Laux}
{\cal L}_{\rm aux}^Q &=&  \bar \psi^Q_n gn\mcdot (A_Q\!+\!A_{\bn}^X\!+\!A_{\bn})
  \frac{\bnslash}{2}( \psi_n^L+\xi_n)
  + \bar \psi^Q_n \big[ n\mcdot {\cal P} + gn\mcdot 
  (A_Q\!+\!A^X_{\bn}\!+\!A_{\bn}) \big] \frac{\bnslash}{2} \psi^Q_n \nn\\
 && + (n\leftrightarrow \bn) + \mbox{h.c.}  \nn\\
 && + \frac{1}{2g^2}\mbox{tr}\left\{
[iD_Q^\mu + gA_Q^\mu\,, iD_Q^\nu + gA_Q^\nu]^2\right\} + 
\frac{1}{\alpha_Q}\mbox{tr} \{[iD_Q^\mu, A_{Q\mu}]^2\}\,.
\end{eqnarray}
where $iD_Q^\mu = \frac12 n^\mu [\bnP + g\bn\mcdot (A_n^X\!+\! A_n)] + \frac12
\bn^\mu [{\cP} + gn\cdot (A^X_\bn \!+\! A_\bn)]$.  The solution of the equations
of motion for these modes are
\begin{eqnarray} \label{eom1}
  && \psi_n^Q = (X_\bn\!-\!1) (\psi_n^L+\xi_n) \,,\qquad
  \psi_\bn^Q=(X_n\!-\!1) (\psi_\bn^L+\xi_\bn) \,,\nn\\[3pt]
 &&  X_\bn^\dagger X_n = W_n^X W_\bn^{X\dagger}\,.
\end{eqnarray}
(In addition to the last equation a constraint on the components $n\mcdot A_Q$
and $\bn\mcdot A_Q$ also comes from the gauge fixing term, but will not be
needed.)  The terms in the auxiliary Lagrangian involving the $p^2\sim
Q^2\lambda$ fields are~\cite{bpssoft}
\begin{eqnarray}\label{Laux2}
{\cal L}_{\rm aux}^X &=&  
  \bar \psi^L_n gn\mcdot (A_{n}^X\!+\!A_{s}) \frac{\bnslash}{2}\xi_n 
  + \bar \psi^L_n \big[ n\mcdot {\cal P} + gn\mcdot (A^X_{n}\!+\!A_{s}) \big] 
  \frac{\bnslash}{2} \psi^L_n  +(n\leftrightarrow \bn)+ \mbox{h.c.} \\
 &+& \frac{1}{2g^2}\mbox{tr}\Big\{ [iD_{nX}^\mu \!+\! gA_{n}^{X\mu}\,, 
  iD_{nX}^\nu \!+\! gA_n^{X\nu}]^2\Big\} + \frac{1}{\alpha_{n}}\mbox{tr} 
  \Big\{[iD_{nX}^\mu, A_{n\mu}^{X}]^2\Big\} + (n\leftrightarrow \bn) \,,\nn
\end{eqnarray}
where $iD_{nX}^\mu =\frac12 n^\mu [\bnP +g\bn\mcdot A_n]+\frac12 \bn^\mu
[n\mcdot{\cP}+gn\mcdot A_s]$. The solutions for these modes are
\begin{eqnarray} \label{eom2}
  \psi_n^L= (S_n^X\!-\!1) \xi_n \,,\qquad 
  S_n^{X\dagger} W_n^X = W_n S_n^\dagger \,,\nn\\
  \psi_\bn^L = (S_\bn^X\!-\! 1)\xi_\bn \,,  \qquad
  S_\bn^{X\dagger} W_\bn^X = W_\bn S_\bn^\dagger \,.
\end{eqnarray}
Together Eqs.~(\ref{eom1}) and (\ref{eom2}) can be used at leading order to
eliminate the fields representing offshell fluctuations with
$p^2\gg(Q\lambda)^2$. For collinear quarks this leads to the rules in
Eq.~(\ref{Sadd}). Note that we did not need to use the gauge fixing term to
resolve the ambiguity in the implicit solution for the $\bn\mcdot A$ and
$n\mcdot A$ auxillary fields.

As an illustration of these results we discuss the soft-collinear factorization
for the production of a $q_n \bar q_\bn$ pair with a large invariant mass $Q^2$.
This process is mediated in the full theory by the electromagnetic current $J=
\bar \psi\Gamma \psi$ ($\Gamma$ a color singlet). This current will match onto a
current in SCET that is built entirely out of onshell fields. Using the results
in this appendix this current can be systematically derived. To start the quark
field in $J$ matches onto $\xi_n$ plus all possible fields which the auxiliary
Lagrangian can create starting from $\xi_n$, so
\begin{eqnarray} \label{J1}
 J\to (\bar \xi_n + \bar \psi^L_n+\bar\psi^Q_n)\, \Gamma\, (\xi_\bn + 
    \psi^L_\bn+\psi^Q_\bn) \,.
\end{eqnarray}
Integrating out the $p^2\sim Q^2$ fluctuations with Eq.~(\ref{eom1}) and
inserting a hard Wilson coefficient $C$ which depends on label operators turns
Eq.~(\ref{J1}) into
\begin{eqnarray}
 (\bar\xi_n+\bar\psi^L_n) X_\bn^\dagger\,C\Gamma\, X_n (\xi_\bn+\psi^L_\bn)=
 (\bar\xi_n+\bar\psi^L_n) W_n^X\,C\Gamma\, W_\bn^{X\dagger}(\xi_\bn+\psi^L_\bn)
 \,.
\end{eqnarray}
To construct the first operator we used the equations of motion for $\psi_n^Q$
and $\psi_\bn^Q$, and in the second operator we used the equation of motion
identity for the gluons in $X_n$ and $X_\bn$.  In a similar fashion we can now
integrate out the $p^2\sim Q^2\lambda$ fluctuations with Eq.~(\ref{eom2}) to
give
\begin{eqnarray}
 \bar\xi_n S_n^{X\dagger} W_n^X\,C\Gamma\, W_\bn^{X\dagger} S_\bn^X \xi_\bn
 = \bar\xi_n W_n S_n^\dagger \,C\Gamma\, S_\bn W_\bn^\dagger \xi_\bn \,. 
\end{eqnarray}
The operator on the right is the final result used in Eq.~(\ref{OnnWS}), and is
soft, collinear, and usoft gauge invariant. It should be obvious from this
example how the equations of motion in Eqs.~(\ref{eom1}) and (\ref{eom2}) can be
used to determine the factorized form of a general leading order operator.


\end{document}